\documentclass[aps,prd,preprint,groupedaddress]{revtex4}

\usepackage{graphicx}
\usepackage{side}
\usepackage{axodraw}

\def\dfrac#1#2{\displaystyle\frac{#1}{#2}}

\newcommand{\p}{\partial}
\newcommand{\kslash}{k\kern-1ex /}
\newcommand{\pslash}{p\kern-1ex /}
\newcommand{\qslash}{q\kern-1ex /}
\newcommand{\lslash}{l\kern-1ex /}
\newcommand{\sslash}{s\kern-1ex /}
\newcommand{\Dslash}{{\cal D}\kern-1.5ex /}

\newcommand{\beqa}{\begin{eqnarray}}
\newcommand{\eeqa}{\end{eqnarray}}

\newcommand{\be}{\begin{equation}}
\newcommand{\ee}{\end{equation}}
\newcommand{\ben}{\begin{eqnarray}}
\newcommand{\een}{\end{eqnarray}}
\newcommand{\nn}{\nonumber}
\def\lsim{\raise0.3ex\hbox{$<$\kern-0.75em\raise-1.1ex\hbox{$\sim$}}}
\def\gsim{\raise0.3ex\hbox{$>$\kern-0.75em\raise-1.1ex\hbox{$\sim$}}}
\def\simgt{\rlap{\lower 3.5 pt\hbox{$\mathchar \sim$}}\raise 1pt \hbox {$>$}}
\def\simlt{\rlap{\lower 3.5 pt\hbox{$\mathchar \sim$}}\raise 1pt \hbox {$<$}}
\newcommand{\cont}{{\rm cont}}
\newcommand{\latt}{{\rm latt}}

\newcommand{\mf}{{\rm MF}}
\newcommand{\msbar}{{\overline {\rm MS}}}

\newcommand{\pos}{{p^*}}
\newcommand{\qos}{{q^*}}
\newcommand{\qoss}{{q^*_s}}
\newcommand{\qosd}{{q^*_d}}
\newcommand{\lo}{{(0)}}

\newcommand{\intlat}{{\int_{-\pi}^{\pi}\frac{d^4 k}{(2\pi)^4}}}

\newcommand{\mpl}{{m_{p2}}}
\newcommand{\mph}{{m_{p1}}}
\newcommand{\mpllo}{{m_{p2}^{(0)}}}
\newcommand{\mphlo}{{m_{p1}^{(0)}}}
\newcommand{\mpllomf}{{{\tilde m_{p2}}^{(0)}}}
\newcommand{\mphlomf}{{{\tilde m_{p1}}^{(0)}}}
\def\ovec{\partial_\mu\hspace{-0.4cm}\raisebox{1.8ex}{$\rightarrow$}}
\def\antivec{\partial_\mu\hspace{-0.4cm}\raisebox{1.8ex}{$\leftarrow$}}

\begin{document}


\title{Perturbative Determination of \\
Mass Dependent $O(a)$ Improvement Coefficients\\
for the Vector and Axial Vector Currents \\
with a Relativistic Heavy Quark Action}


\author{Sinya Aoki$^a$, Yasuhisa Kayaba$^a$ and Yoshinobu Kuramashi$^b$}
\affiliation{$^a$Institute of Physics, University of Tsukuba, 
Tsukuba, Ibaraki 305-8571, Japan \\
$^b$Institute of Particle and Nuclear Studies,
High Energy Accelerator Research Organization(KEK),
Tsukuba, Ibaraki 305-0801, Japan}


\date{\today}

\begin{abstract}
We carry out a perturbative determination of 
mass dependent renormalization factors and 
$O(a)$ improvement coefficients for the vector 
and axial vector currents 
with a relativistic heavy quark action, which 
we have designed to control $m_Qa$ errors by extending the 
on-shell  $O(a)$ improvement program to 
the case of $m_Q \gg \Lambda_{\rm QCD}$ with $m_Q$ the heavy quark mass.
We discuss what kind of improvement operators are required 
for the heavy-heavy and the heavy-light cases
under the condition that the Euclidean rotational symmetry 
is not retained anymore because of the $m_Qa$ corrections.
Our calculation is performed employing the ordinary perturbation theory
with the fictitious gluon mass as an infrared regulator.
We show that all the improvement coefficients are determined 
free from infrared divergences.  
Results of the renormalization factors and the improvement coefficients
are presented as a function of $m_Q a$ 
for various improved gauge actions as well as the plaquette action. 

\end{abstract}


\maketitle


\section{Introduction}
\label{sec:intro}

This paper is the third in a series of publications\cite{akt,param} on
a new relativistic approach which was recently proposed 
from the view point of the on-shell $O(a)$ improvement program. 
The generic quark action,
proposed first in Ref.~\cite{fnal}, 
is given by
\ben
S_q&=&\sum_x\left[ m_0{\bar q}(x)q(x)
+{\bar q}(x)\gamma_0 D_0q(x)
+\nu \sum_i {\bar q}(x)\gamma_i D_i q(x)\right.\nn\\
&&\left.-\frac{r_t a}{2} {\bar q}(x)D_0^2 q(x)
-\frac{r_s a}{2} \sum_i {\bar q}(x)D_i^2 q(x)\right.\nn\\
&&\left.-\frac{ig a}{2}c_E \sum_i {\bar q}(x)\sigma_{0i}F_{0i} q(x)
-\frac{ig a}{4}c_B \sum_{i,j} {\bar q}(x)\sigma_{ij}F_{ij} q(x)
\right].
\een
While we are allowed to choose $r_t=1$, other four parameters
$\nu$, $r_s$, $c_E$ and $c_B$ should be properly adjusted 
as functions of $m_Q a$ and the gauge coupling constant $g$,
in order to achieve the $O(a)$ improvement 
for all on-shell matrix elements.
In Ref.~\cite{param} we determine $\nu$, $r_s$, $c_E$ 
and $c_B$ up to the one-loop level for various improved gauge actions.
We now report on the $O(a)$ improvement of the vector and axial vector
currents at the one-loop level for the relativistic heavy quark action.

In this paper we first make a general discussion about  
what kind of improvement operators 
are required from the symmetries allowed on the lattice, in which
the Euclidean rotational symmetry is violated
because of $m_Q a$ corrections.
We consider both the heavy-heavy and heavy-light cases, where
the light quark is massless for the latter. 
Following Ref.~\cite{param} we evaluate one-loop diagrams
employing the conventional perturbative method
with the use of 
the fictitious gluon mass to regularize the infrared divergence.
In the massless case this method was successfully applied  
to the calculation of the renormalization constants and the
improvement coefficients for the bilinear operators\cite{gmass,csw_m0}.

This paper is organized as follows. 
In Secs.~II and III we determine the renormalization constants and the
improvement coefficients for the vector and axial vector currents
up to one-loop level. The results are presented 
both for the heavy-heavy and heavy-light cases
as a function of $m_Qa$ with various improved gauge actions in addition to
the ordinary plaquette action.
In Sec.IV we explain how to implement the mean field improvement 
for the renormalization factors.
Our conclusions are summarized in Sec.~V. 
Some preliminary results are presented in Ref.~\cite{latt03}.

The quark and gluon actions and their Feynman rules are already
presented in Sec.~II of Ref.~\cite{param}. We employ 
the notations introduced there without further notice 
throughout this paper.
As for the numerical evaluation of the one-loop diagrams relevant for 
the vertex functions of the vector and axial vector currents, 
we employ the same method used in the perturbative determination of 
the improvement parameters for the relativistic heavy quark action,
whose technical details are described in Sec.~III of Ref.~\cite{param}.
The physical quantities are expressed in lattice units and 
the lattice spacing $a$ is suppressed unless necessary.
We take  SU($N_c$) gauge group with the gauge coupling constant $g$.

\section{$O(a)$ improvement of the vector currents}
\label{sec:improve_v}

We consider the on-shell $O(a)$ improvement of the vector currents
both for the heavy-heavy and heavy-light cases.
Without Euclidean space-time rotational symmetry, 
the renormalized operators with the $O(a)$ improvement is 
written as
\ben
V^{\latt,R}_\mu(x)&=&
Z_V^{\latt} \left[
{\bar q(x)} \gamma_\mu Q(x)
-g^2 c_{V_\mu}^+
\partial_\mu^- \{{\bar q(x)} Q(x)\}
-g^2 c_{V_\mu}^- \partial_\mu^+ \{{\bar q(x)} Q(x)\}\right.\nn\\
&&\left.-g^2 c_{V_\mu}^L \{{\vec
\partial_i}{\bar q(x)}\} \gamma_i 
\gamma_\mu Q(x) 
-g^2 c_{V_\mu}^H {\bar q(x)}
\gamma_\mu \gamma_i \{{\vec \partial_i} Q(x)\}+O(g^4)\right]
\label{eq:v_r}
\een
where $Z_{V_\mu}^{\latt}$ and $c_{V_\mu}^{(+,-,H,L)}$ depend on
the quark masses $m_Q$ and $m_q$. 
$\partial_\mu^+$ and $\partial_\mu^-$ are defined as
$\partial_\mu^+ = \ovec + \antivec$ and 
$\partial_\mu^- = \ovec - \antivec$.
For the time component of the vector currents 
we can choose $c^H_{V_0}=c^L_{V_0}=0$
with the aid of equation of motion.
In the case of $m_Q=m_q$ we find $c_{V_\mu}^-=0$ and
$c_{V_\mu}^H=c_{V_\mu}^L$ from the charge conjugation symmetry.
Once the both quark masses are massless, all the improvement coefficients
except $c_{V_\mu}^+$ vanishes.

In this section $Z_{V_\mu}^{\latt}$ 
and $c_{V_\mu}^{(+,-,H,L)}$ are determined at the one-loop level 
as a function of $m_Q a$ both for the heavy-heavy and heavy-light cases.
We employ the relativistic heavy quark action
proposed by the authors\cite{akt} both for the heavy and light quarks. 

\subsection{Determination of the improvement coefficients for the vector currents}

\begin{figure}[t]
\begin{center}

\begin{picture}(220,180)(0,0)
\ArrowLine(120,150)(72,90)
\ArrowLine(72,90)(40,50)
\Vertex(72,90){2}
\ArrowLine(200,50)(168,90)
\ArrowLine(168,90)(120,150)
\Vertex(168,90){2}
\Gluon(72,90)(168,90){-5}{8}
\LongArrow(196,70)(184,85)
\Vertex(120,150){2}
\Text(196,84)[l]{$p$}
\LongArrow(56,85)(44,70)
\Text(42,84)[l]{$q$}
\end{picture}

\end{center}
\caption{One-loop diagrams for the vertex functions. $q$ denotes the outgoing quark momentum and $p$ denotes the incoming quark momentum.}
\label{fig:vtx_1loop}
\end{figure}
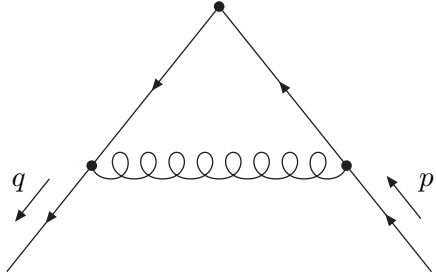

We consider the general form of the off-shell vertex functions
of the vector currents on the lattice at the one-loop level:
\ben
\Lambda_k^{(1)}(p,q,\mph,\mpl) 
&=&\gamma_k F_1^k
+\gamma_k\{\pslash F_2^k+\pslash_s F_3^k\}
+\{\qslash F_4^k+\qslash_s F_5^k\}\gamma_k\nn\\
&&+\qslash\gamma_k\pslash F_6^k
+\qslash\gamma_k\pslash_s F_7^k
+\qslash_s\gamma_k\pslash F_8^k
+\gamma_k\pslash_s\pslash F_9^k
+\qslash\qslash_s\gamma_k F_{10}^k
\nn\\
&&+(p_k+q_k)\left[ G_1^k+\pslash G_2^k
+\qslash G_3^k+\qslash\pslash G_4^k\right] \nn\\
&&+(p_k-q_k)\left[ H_1^k+\pslash H_2^k
+\qslash H_3^k+\qslash\pslash H_4^k\right]+O(a^2)
\label{eq:v_k_s_offsh}
\een
and
\ben
\Lambda_0^{(1)}(p,q,\mph,\mpl) 
&=&\gamma_0 F_1^0
+\gamma_0\pslash F_2^0
+\qslash \gamma_0 F_3^0
+\qslash\gamma_0\pslash F_4^0\nn\\
&&+(p_0+q_0)\left[ G_1^0+\pslash G_2^0
+\qslash G_3^0+\qslash\pslash G_4^0\right]\nn\\
&&+(p_0-q_0)\left[ H_1^0+\pslash H_2^0
+\qslash H_3^0+\qslash\pslash H_4^0\right]+O(a^2)
\label{eq:v_0_s_offsh}
\een  
with
\ben
\pslash&=&\sum_{\alpha=0}^3 p_\alpha \gamma_\alpha, \quad
\qslash =\sum_{\alpha=0}^3 q_\alpha \gamma_\alpha, \quad
\pslash_s =\sum_{i=1}^3 p_i \gamma_i, \quad
\qslash_s =\sum_{i=1}^3 q_i \gamma_i, 
\een
where we assume that the off-shell vertex functions are perturbatively expanded as
\ben
\Lambda_\mu (p,q,\mph,\mpl)=\gamma_\mu+\sum_{i=1} (g^2)^i \Lambda_\mu^{(i)}(p,q,\mph,\mpl).
\een
The vertex functions (\ref{eq:v_k_s_offsh}) 
and (\ref{eq:v_0_s_offsh}) are
defined for the process depicted in Fig.~\ref{fig:vtx_1loop}.
The coefficients $F^\mu$, $G^\mu$, $H^\mu$  are functions of 
$p^2$, $q^2$, $p\cdot q$, $\mph$ and $\mpl$.

Sandwiching (\ref{eq:v_k_s_offsh}) and (\ref{eq:v_0_s_offsh}) 
by the on-shell quark states
$u(p)$ and $\bar u(q)$, which satisfy
$\pslash u(p) = i \mph u(p)$ and $\bar u(q) \qslash = i\mpl \bar u(q)$,
the matrix elements are reduced to 
\ben
&&{\bar u}(q)\Lambda_k^{(1)}(p,q,\mph,\mpl)u(p)\nn\\
&=&{\bar u}(q)\gamma_k u(p)
\left\{F_1^k+i\mph F_2^k+i\mpl F_4^k-\mph\mpl F_6^k\right\} \nn\\
&&+{\bar u}(q)\gamma_k\pslash_s u(p)
\left\{F_3^k+i\mpl F_7^k+i\mph F_9^k\right\} \nn\\
&&+{\bar u}(q)\qslash_s\gamma_k u(p)
\left\{F_5^k+i\mph F_8^k+i\mpl F_{10}^k\right\} \nn\\
&&+(p_k+q_k){\bar u}(q)u(p)
\left\{G_1^k+i\mph G_2^k+i\mpl G_3^k-\mph\mpl G_4^k\right\}\nn \\
&&+(p_k-q_k){\bar u}(q)u(p)
\left\{H_1^k+i\mph H_2^k+i\mpl H_3^k-\mph\mpl H_4^k\right\}
+O(a^2),
\label{eq:v_k_s_onsh}
\een
and
\ben
&&{\bar u}(q)\Lambda_0^{(1)}(p,q,\mph,\mpl)u(p)\nn\\
&=&{\bar u}(q)\gamma_0 u(p)
\left\{F_1^0+i\mph F_2^0+i\mpl F_3^0-\mph\mpl F_4^0\right\} \nn\\
&&+(p_0+q_0){\bar u}(q)u(p)
\left\{G_1^0+i\mph G_2^0+i\mpl G_3^0-\mph\mpl G_4^0\right\}\nn\\
&&+(p_0-q_0){\bar u}(q)u(p)
\left\{H_1^0+i\mph H_2^0+i\mpl H_3^0-\mph\mpl H_4^0\right\}
+O(a^2).
\label{eq:v_0_s_onsh}
\een
For convenience we express the coefficients as
\ben
X_k&=&F_1^k+i\mph F_2^k+i\mpl F_4^k-\mph\mpl F_6^k, \\
Y_k&=&G_1^k+i\mph G_2^k+i\mpl G_3^k-\mph\mpl G_4^k, \\
Z_k&=&H_1^k+i\mph H_2^k+i\mpl H_3^k-\mph\mpl H_4^k, \\
R_k&=&F_3^k+i\mpl F_7^k+i\mph F_9^k, \\
S_k&=&F_5^k+i\mph F_8^k+i\mpl F_{10}^k 
\label{eq:c_k_s_onsh}
\een
and
\ben
X_0&=&F_1^0+i\mph F_2^0+i\mpl F_3^0-\mph\mpl F_4^0, \\
Y_0&=&G_1^0+i\mph G_2^0+i\mpl G_3^0-\mph\mpl G_4^0, \\
Z_0&=&H_1^0+i\mph H_2^0+i\mpl H_3^0-\mph\mpl H_4^0. 
\label{eq:c_0_s_onsh}
\een

Since the above coefficients contain 
both the lattice artifacts and the physical contributions 
which remains in the continuum,
we have to isolate the lattice artifacts in order to determine 
the improvement coefficients in eq.(\ref{eq:v_r}).
The improvement coefficients are given by
\ben
\Delta_{\gamma_k}&=&\left(X_k\right)^\latt-\left(X_k\right)^\cont,
\label{eq:d_v_k}\\
ic_{V_k}^+&=&\left(Y_k\right)^\latt-\left(Y_k\right)^\cont,
\label{eq:+_v_k}\\
ic_{V_k}^-&=&\left(Z_k\right)^\latt-\left(Z_k\right)^\cont,
\label{eq:-_v_k}\\
-ic_{V_k}^L&=&\left(S_k\right)^\latt,\label{eq:l_v_k}\\
ic_{V_k}^H&=&\left(R_k\right)^\latt,\label{eq:h_v_k}\\
\Delta_{\gamma_0}&=&\left(X_0\right)^\latt-\left(X_0\right)^\cont,
\label{eq:d_v_0}\\
ic_{V_0}^+&=&\left(Y_0\right)^\latt-\left(Y_0\right)^\cont,
\label{eq:+_v_0}\\
ic_{V_0}^-&=&\left(Z_0\right)^\latt-\left(Z_0\right)^\cont,
\label{eq:-_v_0}
\een
where the continuum contributions are obtained by 
employing the naive dimensional
regularization (NDR) with the modified minimal subtraction 
scheme ($\msbar$). We have $R_k=S_k=0$ in the continuum 
from the space-time rotational symmetry.
Here it is reminded that in case of $\mpl=\mph$ we obtain
$c_{V_k}^L=c_{V_k}^H$ and $c_{V_k}^-=c_{V_0}^-=0$ 
from the charge conjugation symmetry.

The renormalization factor of the vector currents is obtained by 
\ben
\frac{Z_{V_\mu}^\latt}{Z_{V_\mu}^\cont}
=\sqrt{Z_{Q,\latt}^\lo(\mphlo)}
\sqrt{Z_{q,\latt}^\lo(\mpllo)}\left(1-g^2\Delta_{V_\mu}\right)
\label{eq:z_v}
\een
with
\ben
Z_{Q,\latt}^\lo(\mphlo)&=&\cosh(\mphlo)+r_t \sinh(\mphlo)\\
Z_{q,\latt}^\lo(\mpllo)&=&\cosh(\mpllo)+r_t \sinh(\mpllo)\\
\Delta_{V_\mu}&=&\Delta_{\gamma_\mu}
-\frac{\Delta_{Q}}{2}-\frac{\Delta_{q}}{2},
\een
where $\Delta_{Q,q}$ denote the wave function renormalization factors,
which are already given in Ref.~\cite{param}.
Although we evaluate $Z_{V_\mu}^\cont$ in $\msbar$ scheme with NDR
in this paper, 
the reader may be interested in the value defined in  
$\msbar$ scheme with DRED. The conversion between these two
definitions is easily done by the relation
\ben
Z_{V_\mu}^\cont({\rm NDR})=Z_{V_\mu}^\cont({\rm DRED})-\frac{1}{2} g^2.
\een

Employing a set of special momentum assignments $p=\pos\equiv (p_0=im_{p1}, p_i=0)$ and
$q=\qoss\equiv (q_0=im_{p2}, q_i=0)$ or 
$q=\qosd\equiv (q_0=-im_{p2}, q_i=0)$, where subscripts $s$ and $d$ represent
the scattering and the decay respectively,
we extract
the relevant coefficients $X_k,Y_k,Z_k,R_k,S_k$ for $V_k$ from the off-shell 
vertex function (\ref{eq:v_k_s_offsh}):
\ben
X_k^d&=&\frac{1}{4}{\rm Tr} \left[\Lambda_k^{(1)}(1+\gamma_0)\gamma_k\right]_{p=\pos,q=\qosd},
\label{eq:v_k_x}\\
Y_k^s+Z_k^s&=&\frac{1}{4}{\rm Tr} \left[\frac{\partial}{\partial p_k}
\Lambda_k^{(1)}(1+\gamma_0)-\frac{\partial}{\partial p_i}
\Lambda_k^{(1)}(1+\gamma_0)\gamma_i\gamma_k\right]^{i\ne k}_{p=\pos,q=\qoss},
\label{eq:v_k_ypz}\\
Y_k^s-Z_k^s&=&\frac{1}{4}{\rm Tr} \left[\frac{\partial}{\partial q_k}
\Lambda_k^{(1)}(1+\gamma_0)-\frac{\partial}{\partial q_i}
\Lambda_k^{(1)}(1+\gamma_0)\gamma_k\gamma_i\right]^{i\ne k}_{p=\pos,q=\qoss},
\label{eq:v_k_ymz}\\
X_k^s+2i\mph R_k^s&=&\frac{1}{4}{\rm Tr} 
\left[\Lambda_k^{(1)}\gamma_k(1+\gamma_0)
+2i\mph \frac{\partial}{\partial p_i}
\Lambda_k^{(1)}(1+\gamma_0)\gamma_i\gamma_k\right]^{i\ne k}_{p=\pos,q=\qoss},
\label{eq:v_k_r}\\
X_k^s+2i\mpl S_k^s&=&\frac{1}{4}{\rm Tr} 
\left[\Lambda_k^{(1)}(1+\gamma_0)\gamma_k
+2i\mpl \frac{\partial}{\partial q_i}
\Lambda_k^{(1)}(1+\gamma_0)\gamma_k\gamma_i\right]^{i\ne k}_{p=\pos,q=\qoss},
\label{eq:v_k_s}
\een
where superscripts $s$ and $d$ in $X_k,Y_k,Z_k,R_k,S_k$ represent
their momentum assignments.
On the other hand, we employ 
(\ref{eq:v_0_s_offsh}) to determine
$X_0,Y_0,Z_0$ for $V_0$:
\ben
&&X_0^d=\frac{1}{4}{\rm Tr} \left[\Lambda_0^{(1)}\gamma_0
-i\mph \frac{\partial}{\partial p_k}
\Lambda_0^{(1)}(1+\gamma_0)\gamma_k
+i\mpl \frac{\partial}{\partial q_k}
\Lambda_0^{(1)}(1+\gamma_0)\gamma_k\right]_{p=\pos,q=\qosd},
\label{eq:v_0_x}\\
&&X_0^d+i\mph(Y_0^d+Z_0^d)-i\mpl(Y_0^d-Z_0^d)\nn\\
&&=\frac{1}{4}{\rm Tr} \left[\Lambda_0^{(1)}(1+\gamma_0)
+2i\mpl \frac{\partial}{\partial q_k}
\Lambda_0^{(1)}(1+\gamma_0)\gamma_k\right]_{p=\pos,q=\qosd},
\label{eq:v_0_xyz_d}\\
&&X_0^s+i\mph(Y_0^s+Z_0^s)+i\mpl(Y_0^s-Z_0^s)=
\frac{1}{4}{\rm Tr} \left[\Lambda_0^{(1)}(1+\gamma_0)\right]_{p=\pos,q=\qoss},
\label{eq:v_0_xyz_s}
\een
where we have used the fact that $F^k$, $G^k$ and $H^k$ are functions of
$p^2$, $q^2$ and $p\cdot q$, so that
\ben
&&\left.\frac{\partial F^\mu}{\partial p_i}\right|_{p=\pos,q=\qos}
=\left.\frac{\partial F^\mu}{\partial q_i}\right|_{p=\pos,q=\qos}=0, \\
&&\left.\frac{\partial H^\mu}{\partial p_i}\right|_{p=\pos,q=\qos}
=\left.\frac{\partial H^\mu}{\partial q_i}\right|_{p=\pos,q=\qos}=0, \\
&&\left.\frac{\partial G^\mu}{\partial p_i}\right|_{p=\pos,q=\qos}
=\left.\frac{\partial G^\mu}{\partial q_i}\right|_{p=\pos,q=\qos}=0
\een
with $i=1,2,3$.

Here we briefly explain how to deal with the infrared divergence 
in the above coefficients at the one-loop level.
We basically follow the method employed in Refs.~\cite{kura,param}.
Suppose the vertex function $\Lambda_{k,0}$ at the one-loop level is written as
\ben
\Lambda_{k,0}^{(1)}=\intlat I_{k,0}(k,p,q,\mph,\mpl,\lambda),
\een
where $\lambda$ is the fictitious gluon mass introduced to regularize 
the infrared divergence.
We extract the infrared divergent term as
\ben
&&\intlat I_{k,0}(k,p,q,\mph,\mpl,\lambda)\nn\\
&=&\left.\intlat \left[I_{k,0}(k,p,q,\mph,\mpl,\lambda)
-{\tilde I}_{k,0}(k,p,q,\mph,\mpl,\lambda)\right]\right|_{\lambda\rightarrow 0}\nn\\
&&+\intlat {\tilde I}_{k,0}(k,p,q,\mph,\mpl,\lambda),
\een
where ${\tilde I}_{k,0}(k,p,q,\mph,\mpl,\lambda)$ should have an analytically integrable expression, whose 
infrared behavior is the same as $I_{k,0}(k,p,q,\mph,\mpl,\lambda)$.
For ${\tilde I}_{k,0}(k,p,q,\mph,\mpl,\lambda)$  we employ
\ben
&&{\tilde I}_\mu(k,p,q,\mphlo,\mpllo,\lambda)\nn\\
&=&\theta(\Lambda^2-k^2)C_F
i\gamma_\alpha \frac{1}{i(\qslash+\kslash)+\mpllo}\gamma_\mu\frac{1}{i(\pslash+\kslash)+\mphlo}
i\gamma_\alpha \frac{1}{k^2+\lambda^2},
\een
where $C_F=(N_c^2-1)/(2 N_c)$ denotes the second Casimir of SU($N_c$) group. 
A domain of integration is restricted to
a hypersphere of radius $\Lambda$ $(\le\pi)$ for convenience 
of an analytical integration.

\subsection{Results for the improvement coefficients of the vector currents}

\subsubsection{Heavy-heavy case}

\begin{figure}[b]
\centering{
\hskip -0.0cm
\includegraphics[width=120mm,angle=0]{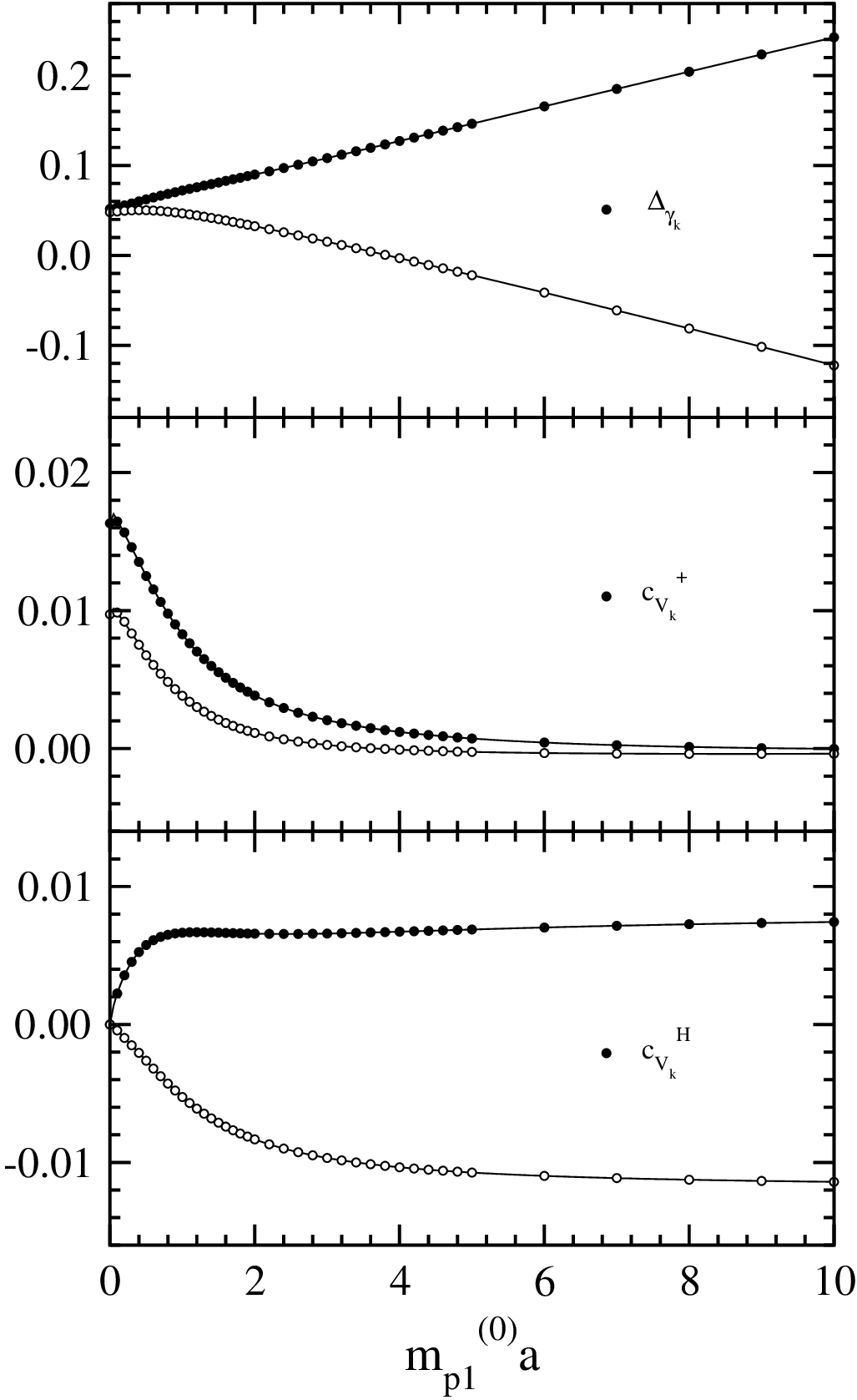}     
}
\caption{$\Delta_{\gamma_k}, c_{V_k}^{(+,H)}$ for heavy-heavy case as 
a function of $\mphlo$. Solid symbols denote the plaquette gauge action
and open ones for the Iwasaki gauge action.}
\label{fig:v_k_hh}
\vspace{8mm}
\end{figure}

\begin{figure}[b]
\centering{
\hskip -0.0cm
\includegraphics[width=120mm,angle=0]{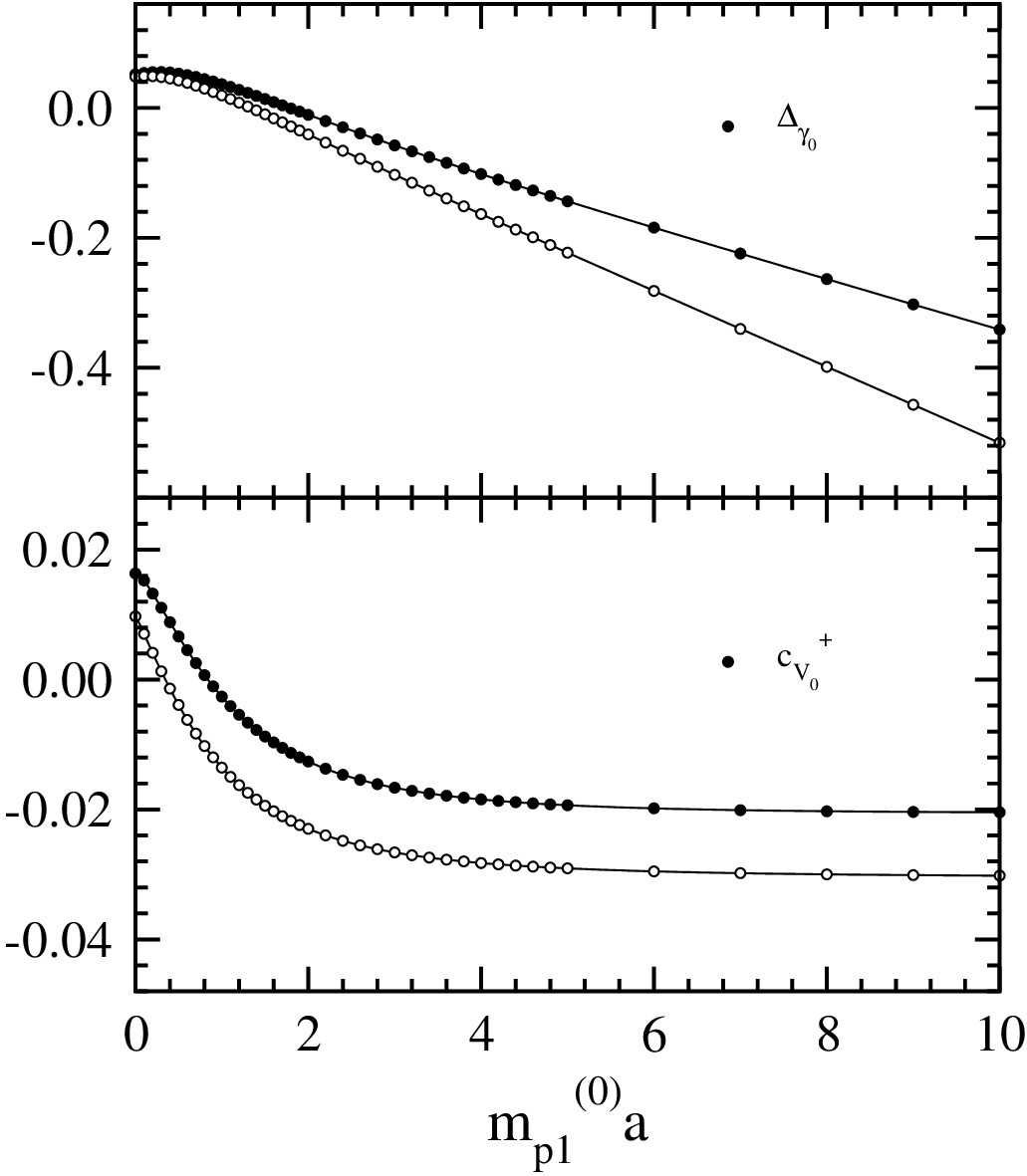}     
}
\caption{$\Delta_{\gamma_0}, c_{V_0}^{+}$ for heavy-heavy case as 
a function of $\mphlo$. Solid symbols denote the plaquette gauge action
and open ones  for the Iwasaki gauge action.}
\label{fig:v_0_hh}
\vspace{8mm}
\end{figure}

\begin{figure}[b]
\centering{
\hskip -0.0cm
\includegraphics[width=120mm,angle=0]{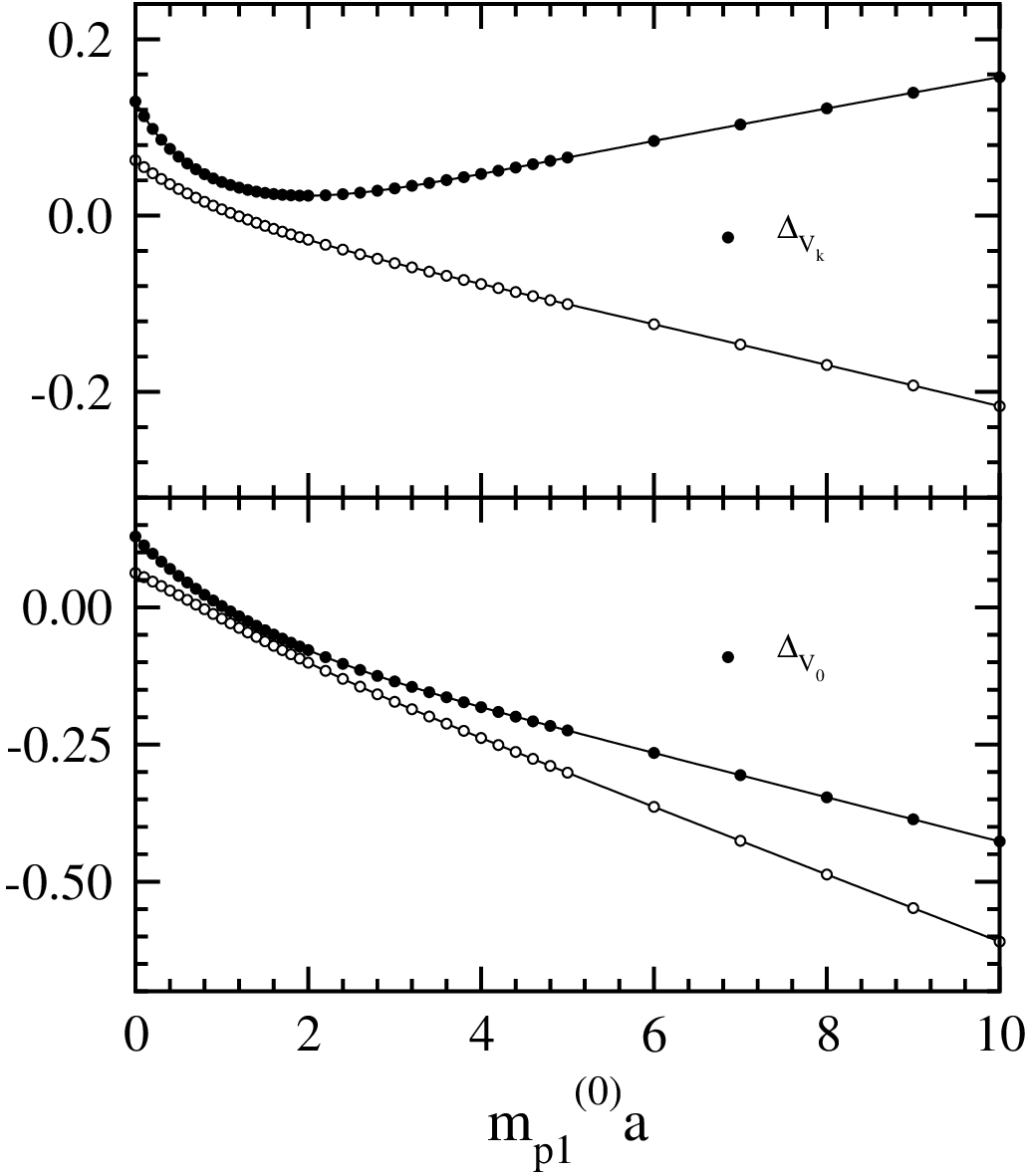}     
}
\caption{$\Delta_{V_\mu}$ for heavy-heavy case as 
a function of $\mphlo$. Solid symbols denote the plaquette gauge action
and open ones  for the Iwasaki gauge action.}
\label{fig:z_v_hh}
\vspace{8mm}
\end{figure}

We obtain $\Delta_{\gamma_k},c_{V_k}^{(+,H)}$ 
and $\Delta_{\gamma_0},c_{V_0}^+$ from 
eqs.(\ref{eq:d_v_k}), (\ref{eq:+_v_k}) 
and (\ref{eq:h_v_k}$-$\ref{eq:+_v_0}) choosing
$\mph=\mpl$, where the charge conjugation symmetry demands
$c_{V_k}^L=c_{V_k}^H$ and $c_{V_k}^-=c_{V_0}^-=0$.
In Figs.~\ref{fig:v_k_hh} and \ref{fig:v_0_hh} 
we plot $\Delta_{\gamma_k},c_{V_k}^{(+,H)}$ and 
$\Delta_{\gamma_0},c_{V_0}^+$, respectively, 
as a function of $\mphlo$ for the plaquette and the Iwasaki gauge actions. 
The $\mphlo$ dependence of $\Delta_{V_\mu}$
in eq.(\ref{eq:z_v}) is also plotted in Fig.~\ref{fig:z_v_hh}.
The solid lines denote the fitting results of the 
interpolation:
\ben
\Delta_{\gamma_k}&=&
\left.\Delta_{\gamma_k}\right|_{\mphlo =0}+
\frac{\sum_{i=1}^5 v_{ni}^{k\gamma}\{\mphlo\}^i}
{1+\sum_{i=1}^5 v_{di}^{k\gamma}\{\mphlo\}^i},
\label{eq:fit_d_v_k_hh}\\
c_{V_k}^+&=&
\left.c_{V_k}^+\right|_{\mphlo =0}+
\frac{\sum_{i=1}^5 v_{ni}^{k+}\{\mphlo\}^i}
{1+\sum_{i=1}^5 v_{di}^{k+}\{\mphlo\}^i},
\label{eq:fit_+_v_k_hh}\\
c_{V_k}^H&=&
\frac{\sum_{i=1}^5 v_{ni}^{kH}\{\mphlo\}^i}
{1+\sum_{i=1}^5 v_{di}^{kH}\{\mphlo\}^i},
\label{eq:fit_h_v_k_hh}\\
\Delta_{V_k}&=&
\left.\Delta_{V_k}\right|_{\mphlo =0}+
\frac{\sum_{i=1}^5 v_{ni}^{kZ}\{\mphlo\}^i}
{1+\sum_{i=1}^5 v_{ni}^{kZ}\{\mphlo\}^i},
\label{eq:fit_z_v_k_hh}\\
\Delta_{\gamma_0}&=&
\left.\Delta_{\gamma_0}\right|_{\mphlo =0}+
\frac{\sum_{i=1}^5 v_{ni}^{0\gamma}\{\mphlo\}^i}
{1+\sum_{i=1}^5 v_{di}^{0\gamma}\{\mphlo\}^i},
\label{eq:fit_d_v_0_hh}\\
c_{V_0}^+&=&
\left.c_{V_0}^+\right|_{\mphlo =0}+
\frac{\sum_{i=1}^5 v_{ni}^{0+}\{\mphlo\}^i}
{1+\sum_{i=1}^5 v_{di}^{0+}\{\mphlo\}^i},
\label{eq:fit_+_v_0_hh}\\
\Delta_{V_0}&=&
\left.\Delta_{V_0}\right|_{\mphlo =0}+
\frac{\sum_{i=1}^5 v_{ni}^{0Z}\{\mphlo\}^i}
{1+\sum_{i=1}^5 v_{di}^{0Z}\{\mphlo\}^i},
\label{eq:fit_z_v_0_hh}
\een
where we assume $c_{V_k}^H=0$ at $\mphlo=0$.
We employ $\Delta_{\gamma_k}=\Delta_{\gamma_0}=$0.05169(plaqette), 
0.04802(Iwasaki\cite{Iwasaki83}), 0.04595(DBW2\cite{dbw2}), 
$c_{V_k}^+=c_{V_0}^+=$0.01633(plaquette), 0.009728(Iwasaki), 0.004884(DBW2) 
and $\Delta_{V_k}=\Delta_{V_0}=$0.1294(plaquette), 0.06279(Iwasaki), 
0.02566(DBW2) at $\mphlo=0$.
The values of the parameters $v_{ni}^{k,0}$ and $v_{di}^{k,0}$ ($i=1,\dots,5$) 
are summarized in Tables~\ref{tab:fit_v_k_hh} and \ref{tab:fit_v_0_hh}.
The relative errors of these interpolations to the data 
are less than a few \% over the range $0\le \mphlo\le 10$.

\subsubsection{Heavy-light case}

\begin{figure}[b]
\centering{
\hskip -0.0cm
\includegraphics[width=120mm,angle=0]{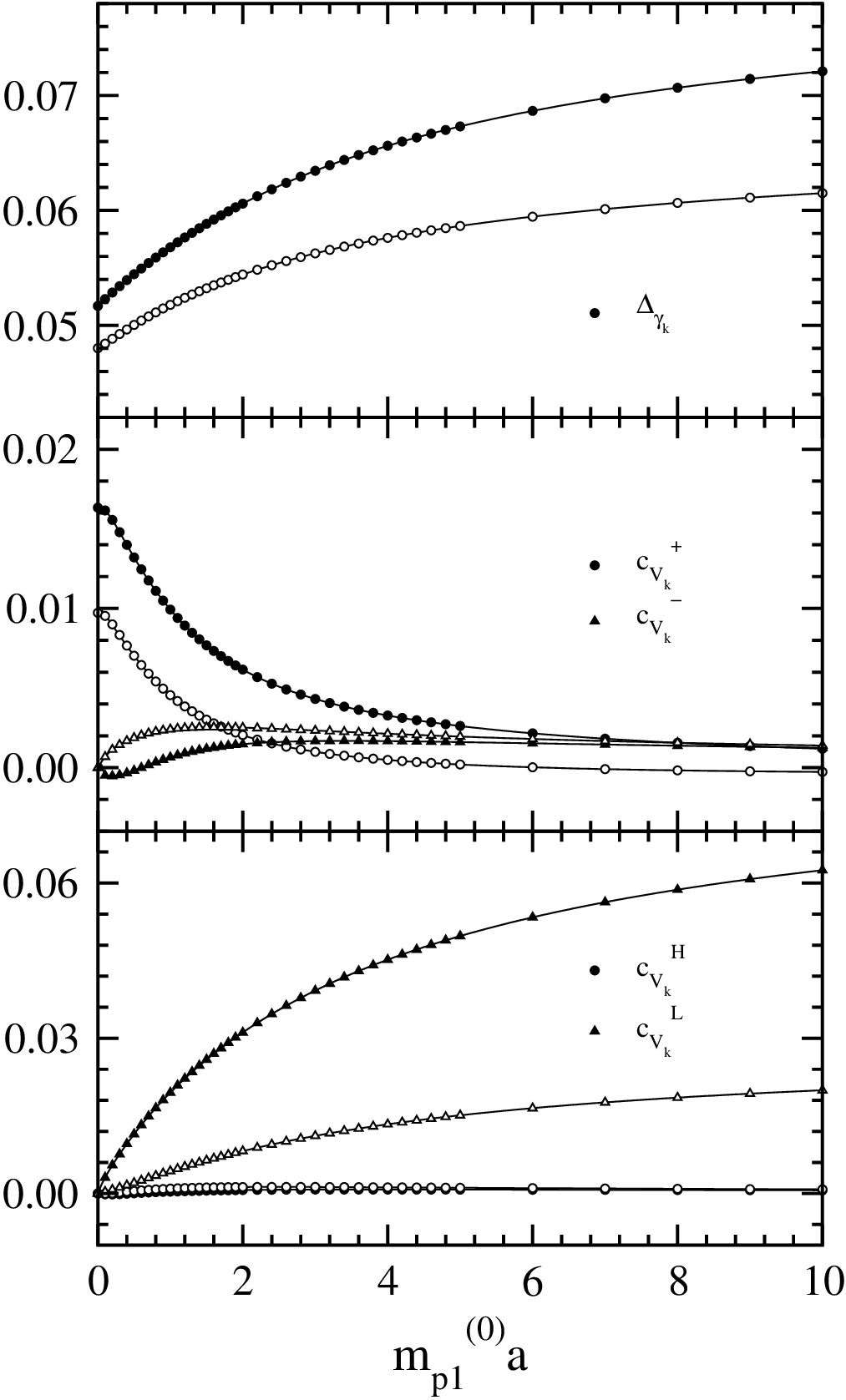}     
}
\caption{$\Delta_{\gamma_k}, c_{V_k}^{(+,-,H,L)}$ for heavy-light case as 
a function of $\mphlo$. Solid symbols denote the plaquette gauge action
and open ones  for the Iwasaki gauge action.}
\label{fig:v_k_hl}
\vspace{8mm}
\end{figure}

\begin{figure}[b]
\centering{
\hskip -0.0cm
\includegraphics[width=120mm,angle=0]{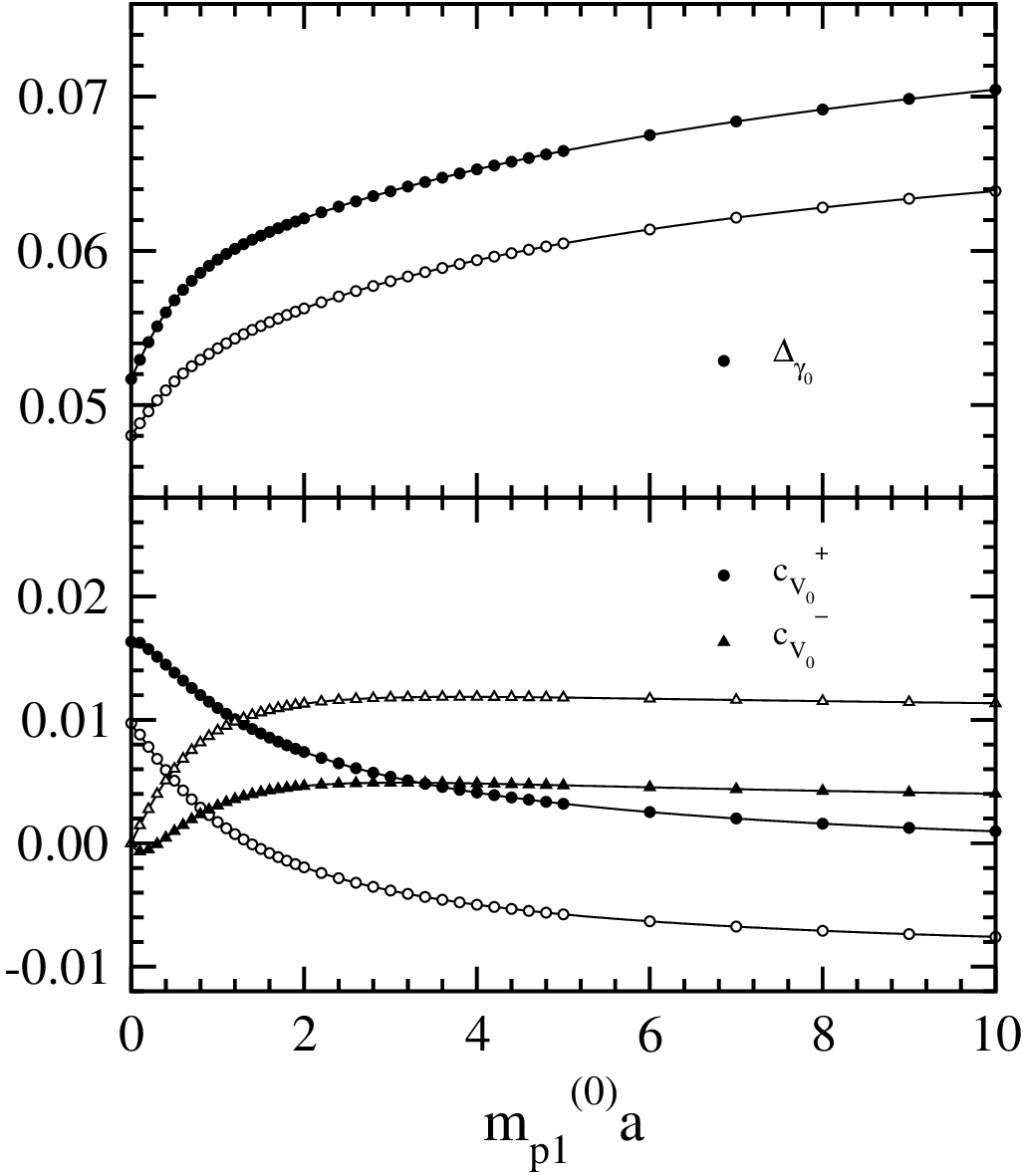}     
}
\caption{$\Delta_{\gamma_0}, c_{V_0}^{(+,-)}$ for heavy-light case as 
a function of $\mphlo$. Solid symbols denote the plaquette gauge action
and open ones  for the Iwasaki gauge action.}
\label{fig:v_0_hl}
\vspace{8mm}
\end{figure}

\begin{figure}[b]
\centering{
\hskip -0.0cm
\includegraphics[width=120mm,angle=0]{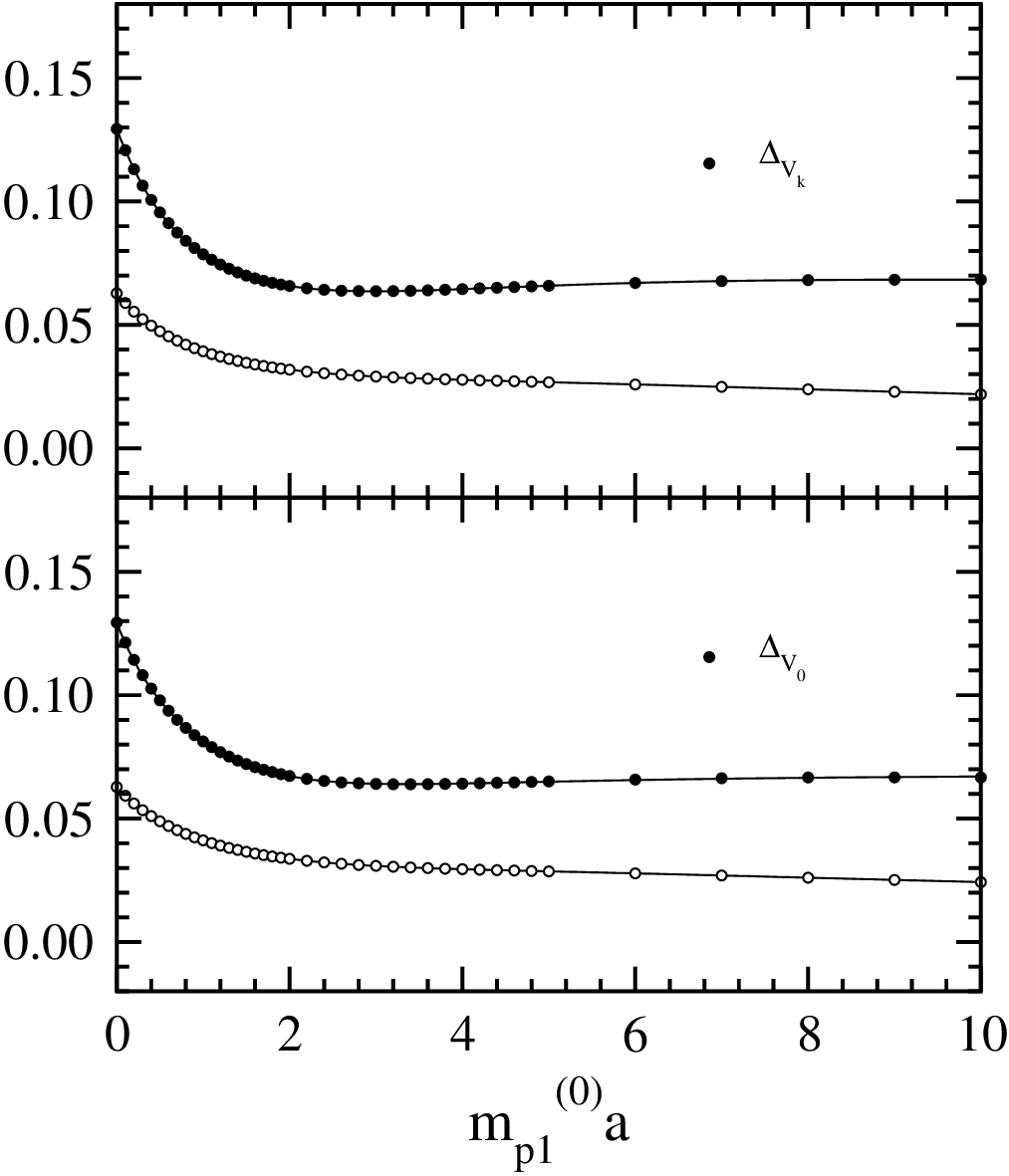}     
}
\caption{$\Delta_{V_\mu}$ for heavy-light case as 
a function of $\mphlo$. Solid symbols denote the plaquette gauge action
and open ones  for the Iwasaki gauge action.}
\label{fig:z_v_hl}
\vspace{8mm}
\end{figure}

We use eqs.(\ref{eq:d_v_k}$-$\ref{eq:-_v_0}) to determine
$\Delta_{\gamma_k}$, $c_{V_k}^{(+,-,H,L)}$ and 
$\Delta_{\gamma_0}$, $c_{V_0}^{(+,-)}$. 
Assuming that the $\mpl a$ corrections are negligible, 
we evaluate the improvement coefficients, 
except $c_{V_k}^{L}$ and $c_{V_0}^{(+,-)}$, as a function of
$\mphlo$ with $\mpllo=0$.
In eqs.(\ref{eq:v_k_s}), (\ref{eq:v_0_xyz_d}), (\ref{eq:v_0_xyz_s}) 
we find that $S_k^s$ and $Y_0^d-Z_0^d$ are not determined
if we set $\mpllo=0$. 
Therefore one should extrapolate data at non-zero 
$\mpllo$ to $\mpllo=0$. We however keep $\mpllo=0.0001$ in our calculation to 
determine $c_{V_k}^{L}$ and $c_{V_0}^{(+,-)}$ since the difference between
the value at $\mpllo=0.0001$ and the one extrapolated to $\mpllo=0$ is
less than
1 \%.
Figures \ref{fig:v_k_hl} and \ref{fig:v_0_hl} show
the $\mphlo$ dependence of 
$\Delta_{\gamma_k}$, $c_{V_k}^{(+,-,H,L)}$ and 
$\Delta_{\gamma_0}$, $c_{V_0}^{(+,-)}$, respectively, 
for the plaquette and the Iwasaki gauge actions.
We also plot $\Delta_{V_\mu}$ in Fig.~\ref{fig:z_v_hl}.
The interpolation denoted by the solid lines are expressed as
\ben
\Delta_{\gamma_k}&=&
\left.\Delta_{\gamma_k}\right|_{\mphlo =0}+
\frac{\sum_{i=1}^5 v_{ni}^{k\gamma}\{\mphlo\}^i}
{1+\sum_{i=1}^5 v_{di}^{k\gamma}\{\mphlo\}^i},
\label{eq:fit_d_v_k_hl}\\
c_{V_k}^{+}&=&
\left.c_{V_k}^{+}\right|_{\mphlo =0}+
\frac{\sum_{i=1}^5 v_{ni}^{k+}\{\mphlo\}^i}
{1+\sum_{i=1}^5 v_{di}^{k+}\{\mphlo\}^i},
\label{eq:fit_+_v_k_hl}\\
c_{V_k}^{(-,H,L)}&=&
\frac{\sum_{i=1}^5 v_{ni}^{k(-,H,L)}\{\mphlo\}^i}
{1+\sum_{i=1}^5 v_{di}^{k(-,H,L)}\{\mphlo\}^i},
\label{eq:fit_-_v_k_hl}\\
\Delta_{V_k}&=&
\left.\Delta_{V_k}\right|_{\mphlo =0}+
\frac{\sum_{i=1}^5 v_{ni}^{kZ}\{\mphlo\}^i}
{1+\sum_{i=1}^5 v_{di}^{kZ}\{\mphlo\}^i},
\label{eq:fit_z_v_k_hl}\nn\\
\Delta_{\gamma_0}&=&
\left.\Delta_{\gamma_0}\right|_{\mphlo =0}+
\frac{\sum_{i=1}^5 v_{ni}^{0\gamma}\{\mphlo\}^i}
{1+\sum_{i=1}^5 v_{di}^{0\gamma}\{\mphlo\}^i},
\label{eq:fit_d_v_0_hl}\\
c_{V_0}^{+}&=&
\left.c_{V_0}^{+}\right|_{\mphlo =0}+
\frac{\sum_{i=1}^5 v_{ni}^{0+}\{\mphlo\}^i}
{1+\sum_{i=1}^5 v_{di}^{0+}\{\mphlo\}^i},
\label{eq:fit_+_v_0_hl}\\
c_{V_0}^{-}&=&
\frac{\sum_{i=1}^5 v_{ni}^{0-}\{\mphlo\}^i}
{1+\sum_{i=1}^5 v_{di}^{0-}\{\mphlo\}^i},
\label{eq:fit_-_v_0_hl}\\
\Delta_{V_0}&=&
\left.\Delta_{V_0}\right|_{\mphlo =0}+
\frac{\sum_{i=1}^5 v_{ni}^{0Z}\{\mphlo\}^i}
{1+\sum_{i=1}^5 v_{di}^{0Z}\{\mphlo\}^i}
\label{eq:fit_z_v_0_hl}
\een
with $v_{ni}^{k,0}$ and $v_{di}^{k,0}$ ($i=1,\dots,5$) given 
in Tables~\ref{tab:fit_v_k_hl} and \ref{tab:fit_v_0_hl}. 
Here we use the constraint that $c_{V_k}^{(-,H,L)}=0$ and $c_{V_0}^{-}=0$ 
at $\mphlo=0$.
The data are well described by these interpolations within 
a few \% errors over the range $0 \le \mphlo\le 10$.

\section{$O(a)$ improvement of the axial vector currents}
\label{sec:improve_a}

Let us turn to the axial vector currents.
The discussion is in parallel with the case of the vector currents.
The renormalized operators with the $O(a)$ improvement is 
given by
\ben
A^{\latt,R}_\mu(x)&=&
Z_A^{\latt} \left[ 
{\bar q(x)} \gamma_\mu\gamma_5 Q(x)
-g^2 c_{A_\mu}^+
\partial_\mu^+ \{{\bar q(x)} \gamma_5 Q(x)\}
-g^2 c_{A_\mu}^- \partial_\mu^- \{{\bar q(x)}
\gamma_5 Q(x)\}\right.\nn\\
&&\left.-g^2 c_{A_\mu}^L \{{\vec
\partial_i}{\bar q(x)}\} \gamma_i 
\gamma_\mu\gamma_5 Q(x) 
-g^2 c_{A_\mu}^H {\bar q(x)}
\gamma_\mu\gamma_5 \gamma_i \{{\vec \partial_i} Q(x)\}+O(g^4)\right]
\label{eq:a_r}\
\een
where we assume that the Euclidean space-time rotational symmetry 
is not retained on the lattice.
The coefficients $Z_{A_\mu}^{\latt}$ and $c_{A_\mu}^{(+,-,H,L)}$ 
are functions of
the quark masses $m_Q$ and $m_q$. 
With the aid of equation of motion
we are allowed to set $c^H_{A_0}=c^L_{A_0}=0$.
In the special case of $m_Q=m_q$, $c_{A_\mu}^-=0$ and
$c_{A_\mu}^H=c_{A_\mu}^L$ is derived from the charge conjugation symmetry.
We also note that all the improvement coefficients
except $c_{A_\mu}^+$ vanishes in the limit of $m_Q=m_q=0$.
We determine $Z_{A_\mu}^{\latt}$ 
and $c_{A_\mu}^{(+,-,H,L)}$ at the one-loop level for
both heavy-heavy and heavy-light cases.

\subsection{Determination of the improvement coefficients 
for the axial vector currents}

The general form of the off-shell vertex functions 
at the one-loop level on the lattice are given by
\ben
\Lambda_{k5}^{(1)}(p,q,\mph,\mpl) 
&=&\gamma_k\gamma_5 F_1^{k5}
+\gamma_k\gamma_5\{\pslash F_2^{k5}+\pslash_s F_3^{k5}\}
+\{\qslash F_4^{k5+}\qslash_s F_5^{k5}\}\gamma_k\gamma_5\nn\\
&&+\qslash\gamma_k\gamma_5\pslash F_6^{k5}
+\qslash\gamma_k\gamma_5\pslash_s F_7^{k5}
+\qslash_s\gamma_k\gamma_5\pslash F_8^{k5}
+\gamma_k\gamma_5\pslash_s\pslash F_9^{k5}
+\qslash\qslash_s\gamma_k\gamma_5 F_{10}^{k5}
\nn\\
&&+(p_k-q_k)\left[ \gamma_5G_1^{k5}+\gamma_5\pslash G_2^{k5}
+\qslash\gamma_5 G_3^{k5}+\qslash\gamma_5\pslash G_4^{k5}\right]\nn\\
&&+(p_k+q_k)\left[ \gamma_5 H_1^{k5}+\gamma_5\pslash H_2^{k5}
+\qslash\gamma_5 H_3^{k5}+\qslash\gamma_5\pslash H_4^{k5}\right]
+O(a^2),
\label{eq:a_k_s_offsh}
\een  
\ben
\Lambda_{05}^{(1)}(p,q,\mph,\mpl) 
&=&\gamma_0\gamma_5 F_1^{05}
+\gamma_0\gamma_5\pslash F_2^{05}
+\qslash \gamma_0\gamma_5 F_3^{05}
+\qslash\gamma_0\gamma_5\pslash F_4^{05}\nn\\
&&+(p_0-q_0)\left[ \gamma_5G_1^{05}+\gamma_5\pslash G_2^{05}
+\qslash\gamma_5 G_3^{05}+\qslash\gamma_5\pslash G_4^{05}\right]\nn\\
&&+(p_0+q_0)\left[ \gamma_5H_1^{05}+\gamma_5\pslash H_2^{05}
+\qslash\gamma_5 H_3^{05}+\qslash\gamma_5\pslash H_4^{05}\right]
+O(a^2),
\label{eq:a_0_s_offsh}
\een  
where 
the coefficients $F^{\mu 5}$, $G^{\mu 5}$, $H^{\mu 5}$ are functions of 
$p^2$, $q^2$, $p\cdot q$, $\mph$ and $\mpl$.

Sandwiching (\ref{eq:a_k_s_offsh}) and (\ref{eq:a_0_s_offsh}) 
by the on-shell quark states
$u(p)$ and $\bar u(q)$, which satisfy
$\pslash u(p) = i \mph u(p)$ and $\bar u(q) \qslash = i\mpl \bar u(q)$,
the matrix elements are reduced to 
\ben
&&{\bar u}(q)\Lambda_{k5}^{(1)}(p,q,\mph,\mpl)u(p)\nn\\
&=&{\bar u}(q)\gamma_k\gamma_5 u(p)
\left\{F_1^{k5}+i\mph F_2^{k5}+i\mpl F_4^{k5}-\mph\mpl F_6^{k5}\right\} \nn\\
&&+{\bar u}(q)\gamma_k\gamma_5\pslash_s u(p)
\left\{F_3^{k5}+i\mpl F_7^{k5}+i\mph F_9^{k5}\right\} \nn\\
&&+{\bar u}(q)\qslash_s\gamma_k\gamma_5 u(p)
\left\{F_5^{k5}+i\mph F_8^{k5}+i\mpl F_{10}^{k5}\right\} \nn\\
&&+(p_k-q_k){\bar u}(q)\gamma_5u(p)
\left\{G_1^{k5}+i\mph G_2^{k5}+i\mpl G_3^{k5}-\mph\mpl G_4^{k5}\right\}\nn\\
&&+(p_k+q_k){\bar u}(q)\gamma_5u(p)
\left\{H_1^{k5}+i\mph H_2^{k5}+i\mpl H_3^{k5}-\mph\mpl H_4^{k5}\right\}
+O(a^2),
\label{eq:a_k_s_onsh}
\een
and
\ben
&&{\bar u}(q)\Lambda_{05}^{(1)}(p,q,\mph,\mpl)u(p)\nn\\
&=&{\bar u}(q)\gamma_0\gamma_5 u(p)
\left\{F_1^{05}+i\mph F_2^{05}+i\mpl F_3^{05}-\mph\mpl F_4^{05}\right\} \nn\\
&&+(p_0-q_0){\bar u}(q)u(p)
\left\{G_1^{05}+i\mph G_2^{05}+i\mpl G_3^{05}-\mph\mpl G_4^{05}\right\}\nn\\
&&+(p_0+q_0){\bar u}(q)\gamma_5u(p)
\left\{H_1^{05}+i\mph H_2^{05}+i\mpl H_3^{05}-\mph\mpl H_4^{05}\right\}
+O(a^2),
\label{eq:a_0_s_onsh}
\een
where the coefficients are summarized as
\ben
X_{k5}&=&F_1^{k5}+i\mph F_2^{k5}+i\mpl F_4^{k5}-\mph\mpl F_6^{k5}, \\
Y_{k5}&=&G_1^{k5}+i\mph G_2^{k5}+i\mpl G_3^{k5}-\mph\mpl G_4^{k5}, \\
Z_{k5}&=&H_1^{k5}+i\mph H_2^{k5}+i\mpl H_3^{k5}-\mph\mpl H_4^{k5}, \\
R_{k5}&=&F_3^{k5}+i\mpl F_7^{k5}+i\mph F_9^{k5}, \\
S_{k5}&=&F_5^{k5}+i\mph F_8^{k5}+i\mpl F_{10}^{k5}
\label{eq:c_k5_s_onsh}
\een
and
\ben
X_{05}&=&F_1^{05}+i\mph F_2^{05}+i\mpl F_3^{05}-\mph\mpl F_4^{05}, \\
Y_{05}&=&G_1^{05}+i\mph G_2^{05}+i\mpl G_3^{05}-\mph\mpl G_4^{05}, \\
Z_{05}&=&H_1^{05}+i\mph H_2^{05}+i\mpl H_3^{05}-\mph\mpl H_4^{05}. 
\label{eq:c_05_s_onsh}
\een

In terms of these coefficients the improvement coefficients in eq.(\ref{eq:a_r}) are given by
\ben
\Delta_{\gamma_k\gamma_5}&=&\left(X_{k5}\right)^\latt
-\left(X_{k5}\right)^\cont,
\label{eq:d_a_k}\\
ic_{A_k}^+&=&\left(Y_{k5}\right)^\latt-\left(Y_{k5}\right)^\cont,
\label{eq:+_a_k}\\
ic_{A_k}^-&=&\left(Z_{k5}\right)^\latt-\left(Z_{k5}\right)^\cont,
\label{eq:-_a_k}\\
-ic_{A_k}^L&=&\left(S_{k5}\right)^\latt,
\label{eq:l_a_k}\\
ic_{A_k}^H&=&\left(R_{k5}\right)^\latt,
\label{eq:h_a_k}\\
\Delta_{\gamma_0\gamma_5}&=&\left(X_{05}\right)^\latt-\left(X_{05}\right)^\cont,
\label{eq:d_a_0}\\
ic_{A_0}^+&=&\left(Y_{05}\right)^\latt-\left(Y_{05}\right)^\cont,
\label{eq:+_a_0}\\
ic_{A_0}^-&=&\left(Z_{05}\right)^\latt-\left(Z_{05}\right)^\cont,
\label{eq:-_a_0}
\een
where we calculate the continuum contributions
employing the $\msbar$ scheme with NDR.
It should be noted that $R_{k5}=S_{k5}=0$ 
in the continuum from the space-time rotational symmetry and 
$c_{A_k}^L=c_{A_k}^H$ and $c_{A_k}^-=c_{A_0}^-=0$ 
for $\mph=\mpl$ from the charge conjugation symmetry.

Combining $\Delta_{\gamma_\mu\gamma_5}$ and the wave function 
renormalization factors, we obtain
the renormalization factor of the axial vector currents:
\ben
\frac{Z_{A_\mu}^\latt}{Z_{A_\mu}^\cont}
=\sqrt{Z_{Q,\latt}^\lo(\mphlo)}
\sqrt{Z_{q,\latt}^\lo(\mpllo)}\left(1-g^2\Delta_{A_\mu}\right)
\label{eq:z_a}
\een
with
\ben
Z_{Q,\latt}^\lo(\mphlo)&=&\cosh(\mphlo)+r_t \sinh(\mphlo)\\
Z_{q,\latt}^\lo(\mpllo)&=&\cosh(\mpllo)+r_t \sinh(\mpllo)\\
\Delta_{A_\mu}&=&\Delta_{\gamma_\mu\gamma_5}
-\frac{\Delta_{Q}}{2}-\frac{\Delta_{q}}{2},
\een
where $\Delta_{Q,q}$ are found in Ref.~\cite{param}.
For convenience we give the relation for 
$Z_{A_\mu}^\cont$ between NDR and DRED in $\msbar$ scheme:
\ben
Z_{A_\mu}^\cont({\rm NDR})=Z_{A_\mu}^\cont({\rm DRED})-\frac{1}{2}g^2.
\een

The relevant coefficients $X_{k5},Y_{k5},Z_{k5},R_{k5},S_{k5}$ for $A_k$ are determined from the
off-shell vertex function (\ref{eq:a_k_s_offsh}):
\ben
X_{k5}^s&=&\frac{1}{4}{\rm Tr} \left[\Lambda_{k5}^{(1)}(1+\gamma_0)\gamma_5\gamma_k\right]_{p=\pos,q=\qoss},
\label{eq:a_k_x}\\
Y_{k5}^d+Z_{k5}^d&=&\frac{1}{4}{\rm Tr} \left[\frac{\partial}{\partial p_k}
\Lambda_{k5}^{(1)}(1+\gamma_0)\gamma_5-\frac{\partial}{\partial p_i}
\Lambda_{k5}^{(1)}(1+\gamma_0)\gamma_i\gamma_k\gamma_5\right]^{i\ne k}_{p=\pos,q=\qosd},\label{eq:a_k_ypz}\\
Y_{k5}^d-Z_{k5}^d&=&\frac{1}{4}{\rm Tr} \left[-\frac{\partial}{\partial q_k}
\Lambda_{k5}^{(1)}(1+\gamma_0)\gamma_5+\frac{\partial}{\partial q_i}
\Lambda_{k5}^{(1)}(1+\gamma_0)\gamma_k\gamma_i\gamma_5\right]^{i\ne k}_{p=\pos,q=\qosd},\label{eq:a_k_ymz}\\
X_{k5}^d+2i\mph R_{k5}^d&=&\frac{1}{4}{\rm Tr} 
\left[\Lambda_{k5}^{(1)}\gamma_5\gamma_k(1-\gamma_0)
+2i\mph \frac{\partial}{\partial p_i}
\Lambda_k^{(1)}(1+\gamma_0)\gamma_i\gamma_5\gamma_k\right]^{i\ne k}_{p=\pos,q=\qosd},\label{eq:a_k_r}\\
X_{k5}^d+2i\mpl S_{k5}^d&=&\frac{1}{4}{\rm Tr} 
\left[\Lambda_k^{(1)}(1+\gamma_0)\gamma_5\gamma_k
+2i\mpl \frac{\partial}{\partial q_i}
\Lambda_k^{(1)}(1+\gamma_0)\gamma_5\gamma_k\gamma_i\right]^{i\ne k}_{p=\pos,q=\qosd},\label{eq:a_k_s}
\een
where $\pos\equiv (p_0=im_{p1}, p_i=0)$ and
$\qoss\equiv (q_0=im_{p2}, q_i=0)$ or $\qosd\equiv (q_0=-im_{p2}, q_i=0)$.
We obtain $X_{05},Y_{05},Z_{05}$ for $A_0$ from eq.(\ref{eq:a_0_s_offsh}):
\ben
&&X_{05}^s\nn\\
&&=\frac{1}{4}{\rm Tr} \left[\Lambda_{05}^{(1)}\gamma_5\gamma_0
+i\mph \frac{\partial}{\partial p_k}
\Lambda_{05}^{(1)}(1+\gamma_0)\gamma_k\gamma_5
+i\mpl \frac{\partial}{\partial q_k}
\Lambda_{05}^{(1)}(1+\gamma_0)\gamma_k\gamma_5\right]_{p=\pos,q=\qoss},
\label{eq:a_0_x}\\
&&X_{05}^s+i\mph(Y_{05}^s+Z_{05}^s)-i\mpl(Y_{05}^s-Z_{05}^s)\nn\\
&&=\frac{1}{4}{\rm Tr} \left[\Lambda_{05}^{(1)}\gamma_5(1+\gamma_0)
+2i\mph \frac{\partial}{\partial p_k}
\Lambda_{05}^{(1)}(1+\gamma_0)\gamma_k\gamma_5\right]_{p=\pos,q=\qoss},
\label{eq:a_0_xyz_s}\\
&&X_{05}^d-i\mph(Y_{05}^d+Z_{05}^d)-i\mpl(Y_{05}^d-Z_{05}^d)=
-\frac{1}{4}{\rm Tr} \left[\Lambda_{05}^{(1)}(1+\gamma_0)\gamma_5\right]_{p=\pos,q=\qosd},
\label{eq:a_0_xyz_d}
\een
where $F^k$, $G^k$ and $H^k$ are functions of
$p^2$, $q^2$ and $p\cdot q$ resulting in
\ben
&&\left.\frac{\partial F^{\mu 5}}{\partial p_i}\right|_{p=\pos,q=\qos}
=\left.\frac{\partial F^{\mu 5}}{\partial q_i}\right|_{p=\pos,q=\qos}=0, \\
&&\left.\frac{\partial H^{\mu 5}}{\partial p_i}\right|_{p=\pos,q=\qos}
=\left.\frac{\partial H^{\mu 5}}{\partial q_i}\right|_{p=\pos,q=\qos}=0, \\
&&\left.\frac{\partial G^{\mu 5}}{\partial p_i}\right|_{p=\pos,q=\qos}
=\left.\frac{\partial G^{\mu 5}}{\partial q_i}\right|_{p=\pos,q=\qos}=0 
\een
with $i=1,2,3$.

As for the counterterm to isolate the infrared divergence in the above coefficients at the one-loop level
we employ
\ben
&&{\tilde I}_{\mu 5}(k,p,q,\mphlo,\mpllo,\lambda)\nn\\
&=&\theta(\Lambda^2-k^2)C_F
i\gamma_\alpha \frac{1}{i(\qslash+\kslash)+\mpllo}\gamma_\mu\gamma_5\frac{1}{i(\pslash+\kslash)+\mphlo}
i\gamma_\alpha \frac{1}{k^2+\lambda^2}
\een
with  a cutoff $\Lambda$ $(\le\pi)$.

\subsection{Results for the improvement coefficients 
of the axial vector currents}

\subsubsection{Heavy-heavy case}

\begin{figure}[b]
\centering{
\hskip -0.0cm
\includegraphics[width=120mm,angle=0]{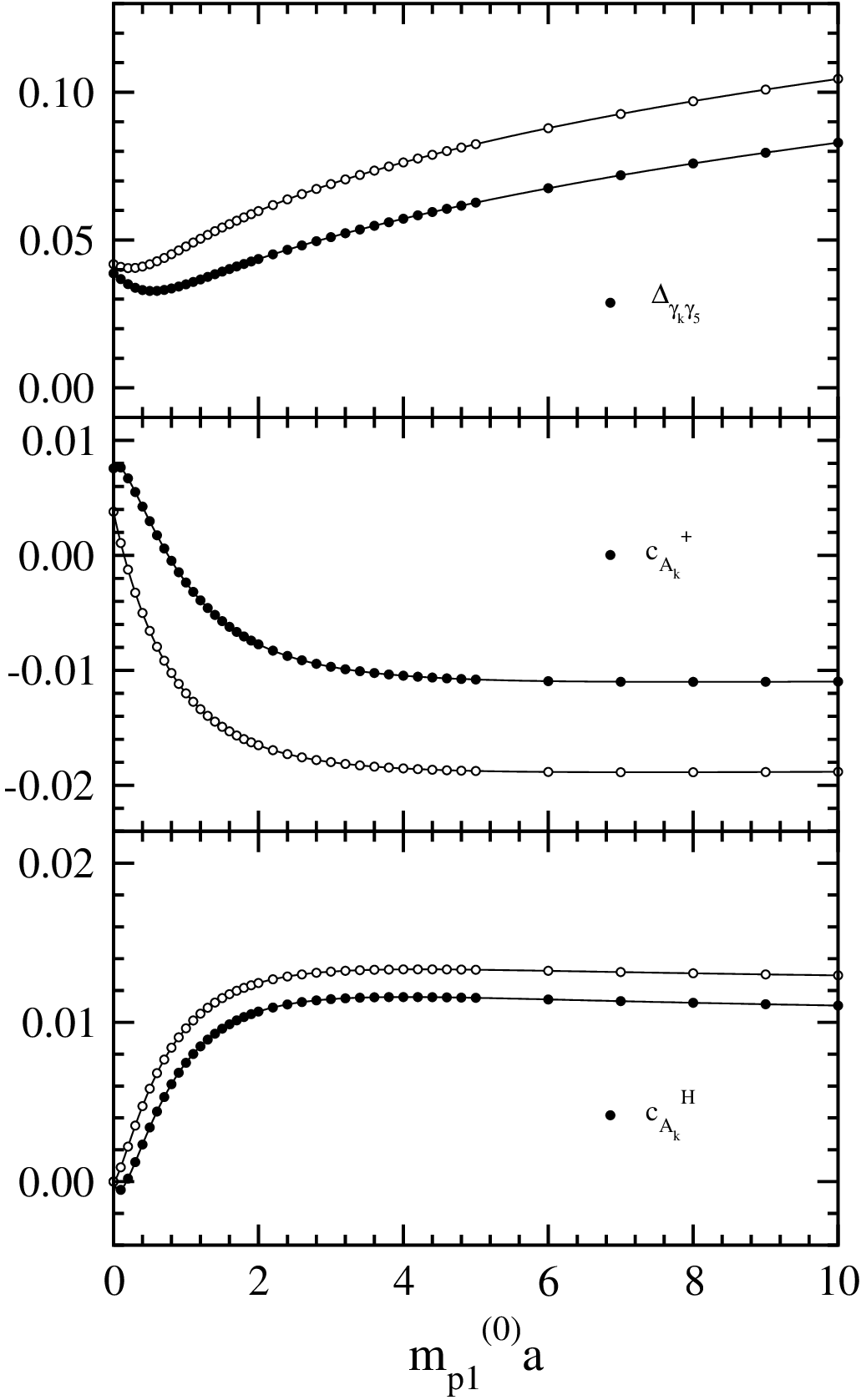}     
}
\caption{$\Delta_{\gamma_k\gamma_5}, c_{A_k}^{(+,H)}$ for heavy-heavy case as 
a function of $\mphlo$. Solid symbols denote the plaquette gauge action
and open ones  for the Iwasaki gauge action.}
\label{fig:a_k_hh}
\vspace{8mm}
\end{figure}

\begin{figure}[b]
\centering{
\hskip -0.0cm
\includegraphics[width=120mm,angle=0]{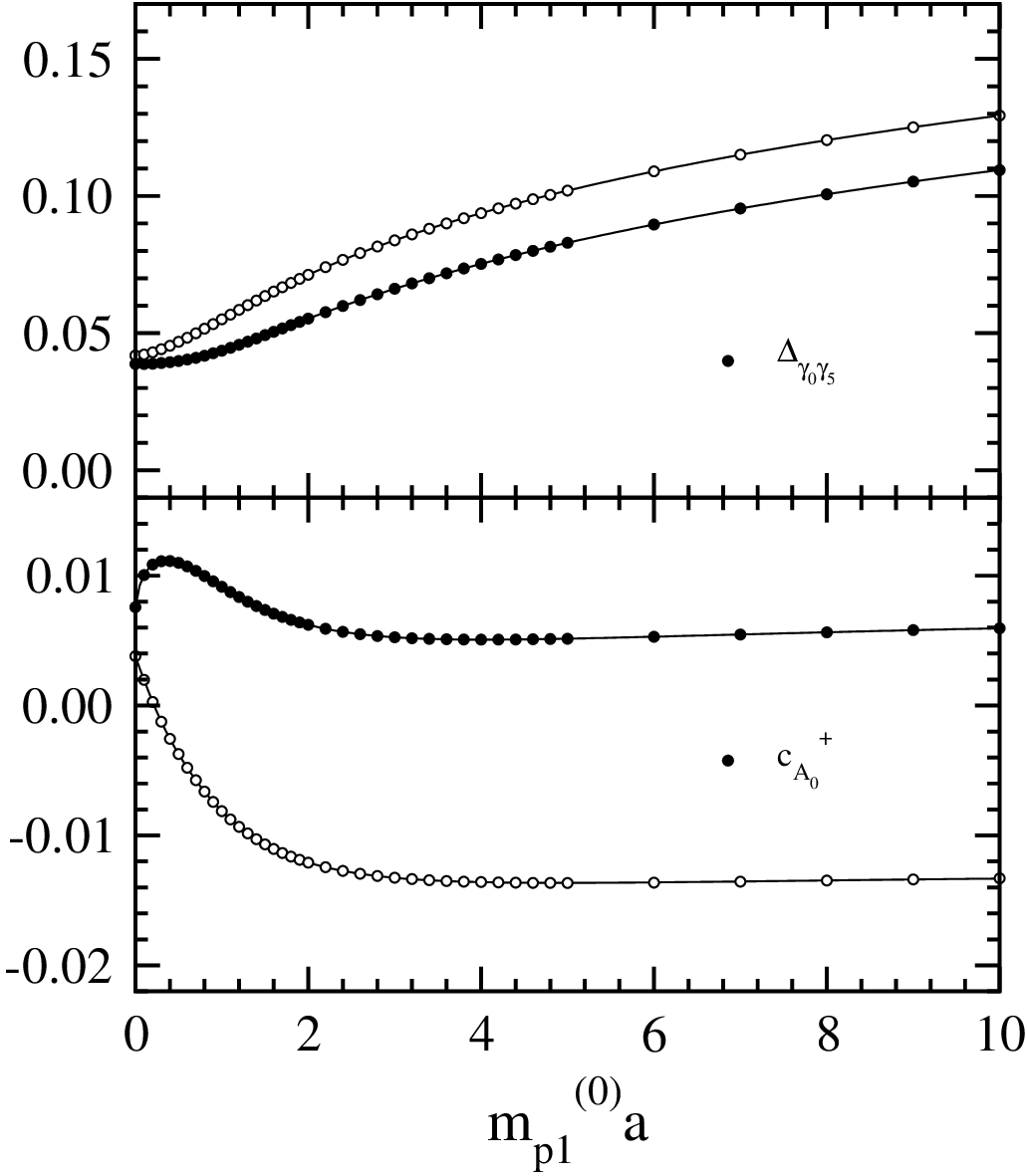}     
}
\caption{$\Delta_{\gamma_0\gamma_5}, c_{A_0}^{+}$ for heavy-heavy case as 
a function of $\mphlo$. Solid symbols denote the plaquette gauge action
and open ones  for the Iwasaki gauge action.}
\label{fig:a_0_hh}
\vspace{8mm}
\end{figure}

\begin{figure}[b]
\centering{
\hskip -0.0cm
\includegraphics[width=120mm,angle=0]{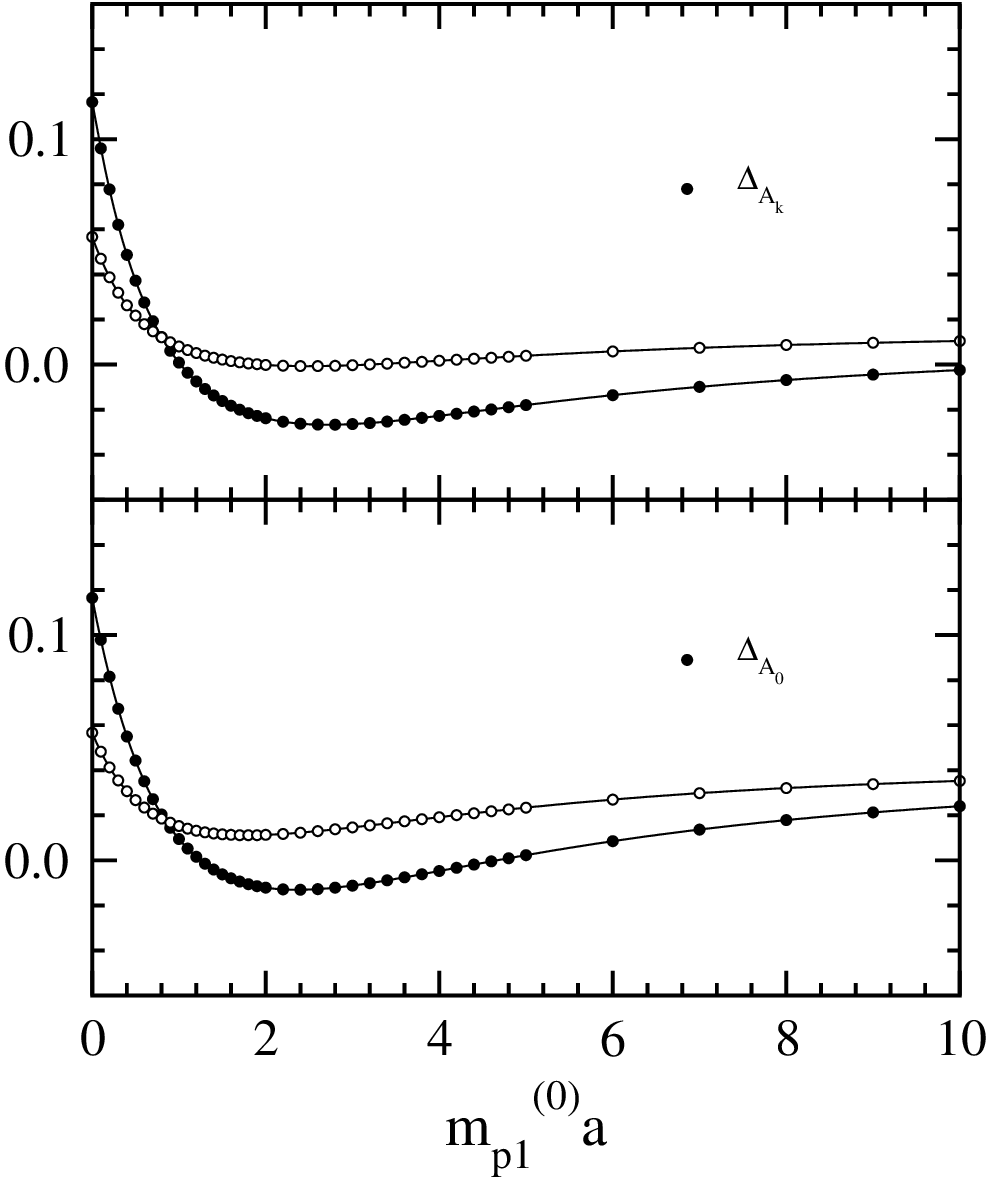}     
}
\caption{$\Delta_{A_\mu}$ for heavy-heavy case as 
a function of $\mphlo$. Solid symbols denote the plaquette gauge action
and open ones  for the Iwasaki gauge action.}
\label{fig:z_a_hh}
\vspace{8mm}
\end{figure}

With the choice of $\mph=\mpl$
the improvement coefficients $\Delta_{\gamma_k\gamma_5},c_{A_k}^{(+,H)}$ 
and $\Delta_{\gamma_0\gamma_5},c_{A_0}^+$ are determined from 
eqs.(\ref{eq:d_a_k}), (\ref{eq:+_a_k}) 
and (\ref{eq:h_a_k}$-$\ref{eq:+_a_0}). 
It is noted that $c_{A_k}^L=c_{A_k}^H$ and $c_{A_k}^-=c_{A_0}^-=0$ from 
the charge conjugation symmetry.
The quark mass dependences of $\Delta_{\gamma_k\gamma_5},c_{A_k}^{(+,H)}$ and 
$\Delta_{\gamma_0\gamma_5},c_{A_0}^+$ are shown in 
Figs.~\ref{fig:a_k_hh} and \ref{fig:a_0_hh}, respectively, 
employing the plaquette and the Iwasaki gauge actions. 
We also give the $\mphlo$ dependence of
$\Delta_{A_\mu}$ in Fig.~\ref{fig:z_a_hh}.
The solid lines denote the interpolation with the use of the following functions:
\ben
\Delta_{\gamma_k\gamma_5}&=&
\left.\Delta_{\gamma_k\gamma_5}\right|_{\mphlo =0}+
\frac{\sum_{i=1}^5 a_{ni}^{k\gamma}\{\mphlo\}^i}
{1+\sum_{i=1}^5 a_{di}^{k\gamma}\{\mphlo\}^i},
\label{eq:fit_d_a_k_hh}\\
c_{A_k}^+&=&
\left.c_{A_k}^+\right|_{\mphlo =0}+
\frac{\sum_{i=1}^5 a_{ni}^{k+}\{\mphlo\}^i}
{1+\sum_{i=1}^5 a_{di}^{k+}\{\mphlo\}^i},
\label{eq:fit_+_a_k_hh}\\
c_{A_k}^H&=&
\frac{\sum_{i=1}^5 a_{ni}^{kH}\{\mphlo\}^i}
{1+\sum_{i=1}^5 a_{di}^{k+}\{\mphlo\}^i},
\label{eq:fit_h_a_k_hh}\\
\Delta_{A_k}&=&
\left.\Delta_{A_k}\right|_{\mphlo =0}+
\frac{\sum_{i=1}^5 a_{ni}^{kZ}\{\mphlo\}^i}
{1+\sum_{i=1}^5 a_{ni}^{kZ}\{\mphlo\}^i},
\label{eq:fit_z_a_k_hh}\\
\Delta_{\gamma_0\gamma_5}&=&
\left.\Delta_{\gamma_0\gamma_5}\right|_{\mphlo =0}+
\frac{\sum_{i=1}^5 a_{ni}^{0\gamma}\{\mphlo\}^i}
{1+\sum_{i=1}^5 a_{di}^{0\gamma}\{\mphlo\}^i},
\label{eq:fit_d_a_0_hh}\\
c_{A_0}^+&=&
\left.c_{A_0}^+\right|_{\mphlo =0}+
\frac{\sum_{i=1}^5 a_{ni}^{0+}\{\mphlo\}^i}
{1+\sum_{i=1}^5 a_{di}^{0+}\{\mphlo\}^i},
\label{eq:fit_+_a_0_hh}\\
\Delta_{A_0}&=&
\left.\Delta_{A_0}\right|_{\mphlo =0}+
\frac{\sum_{i=1}^5 a_{ni}^{0Z}\{\mphlo\}^i}
{1+\sum_{i=1}^5 a_{di}^{0Z}\{\mphlo\}^i},
\label{eq:fit_z_a_0_hh}
\een
with $a_{ni}^{k,0}$ and $a_{di}^{k,0}$ ($i=1,\dots,5$) given 
in Tables~\ref{tab:fit_a_k_hh} and \ref{tab:fit_a_0_hh}.
The relative errors of these interpolations are 
a few \% over the range $0\le \mphlo\le 10$.
The massless values are
$\Delta_{\gamma_k\gamma_5}=\Delta_{\gamma_0\gamma_5}=$0.03873(plaqette),
0.04184(Iwasaki), 0.04359(DBW2),
$c_{A_k}^+=c_{A_0}^+=$0.007574(plaquette), 0.003801(Iwasaki),
0.001492(DBW2) and 
$\Delta_{A_k}=\Delta_{A_0}=$0.1165(plaquette), 0.05663(Iwasaki),
0.02330(DBW2). 
The coefficient $c_{A_k}^H$ should vanish at $\mphlo=0$.

\subsubsection{Heavy-light case}

The set of improvement coefficients are determined from
eqs.(\ref{eq:d_a_k}$-$\ref{eq:-_a_0}).
For the same reason before
we keep $\mpllo=0.0001$ for the determination of $c_{A_k}^{L}$ and $c_{A_0}^{(+,-)}$,
while the vanishing light quark mass $\mpllo=0$ is employed for 
other coefficients.  

\begin{figure}[b]
\centering{
\hskip -0.0cm
\includegraphics[width=120mm,angle=0]{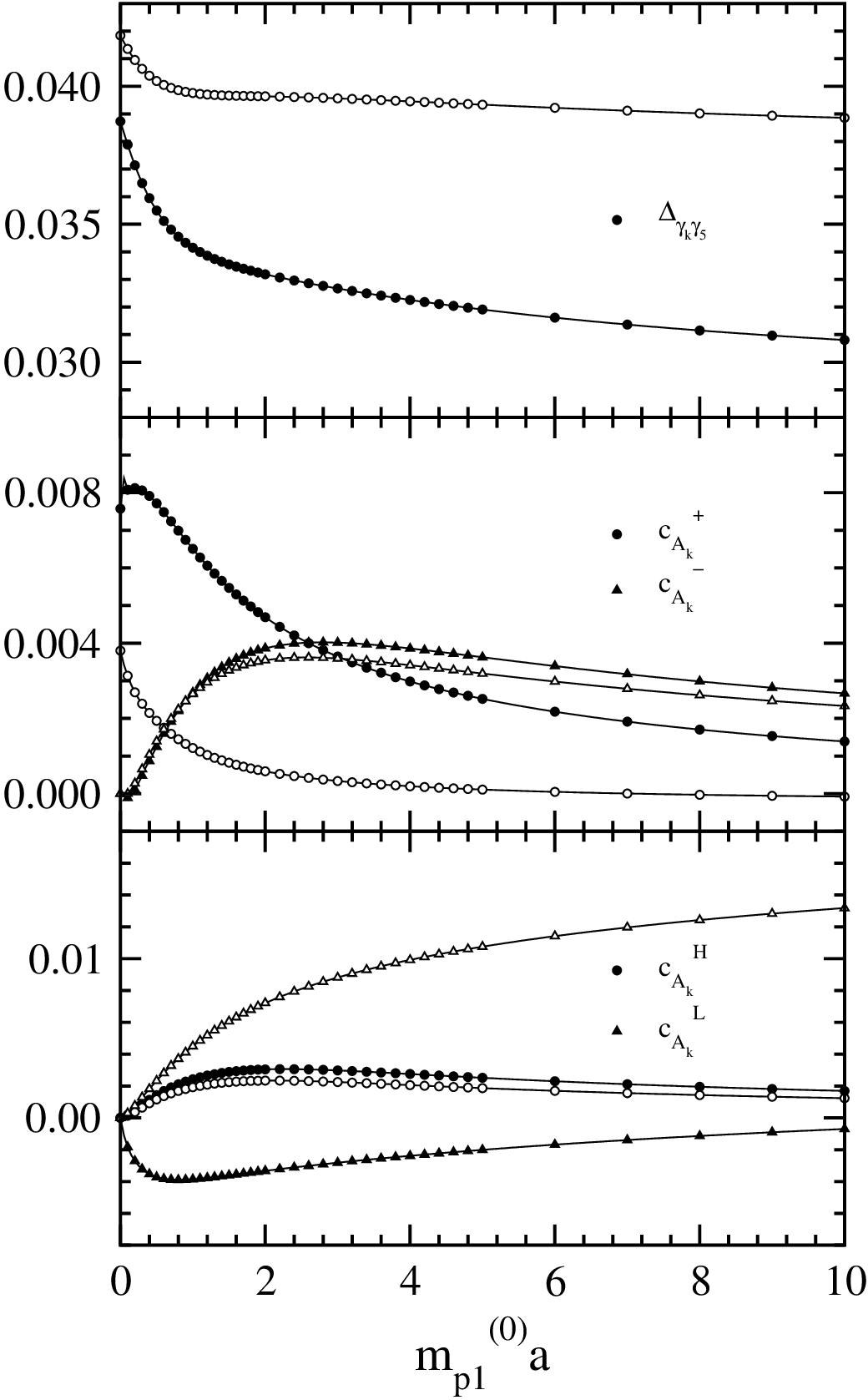}     
}
\caption{$\Delta_{\gamma_k\gamma_5}, c_{A_k}^{(+,-,H,L)}$ 
for heavy-light case as 
a function of $\mphlo$. Solid symbols denote the plaquette gauge action
and open ones for the Iwasaki gauge action.}
\label{fig:a_k_hl}
\vspace{8mm}
\end{figure}

\begin{figure}[b]
\centering{
\hskip -0.0cm
\includegraphics[width=120mm,angle=0]{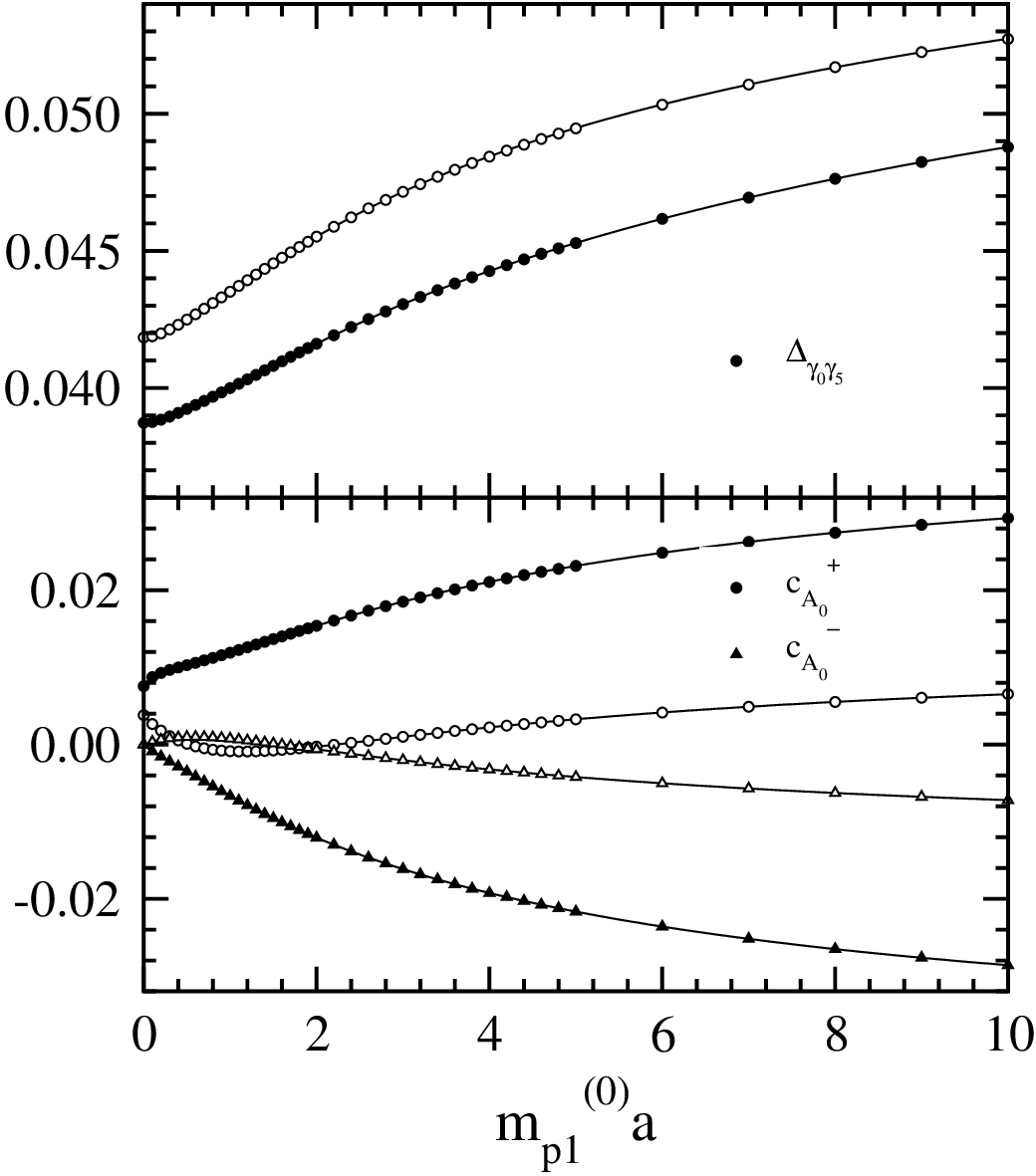}     
}
\caption{$\Delta_{\gamma_0\gamma_5}, c_{A_0}^{(+,-)}$ for heavy-light case as 
a function of $\mphlo$. Solid symbols denote the plaquette gauge action
and open ones for the Iwasaki gauge action.}
\label{fig:a_0_hl}
\vspace{8mm}
\end{figure}

\begin{figure}[b]
\centering{
\hskip -0.0cm
\includegraphics[width=120mm,angle=0]{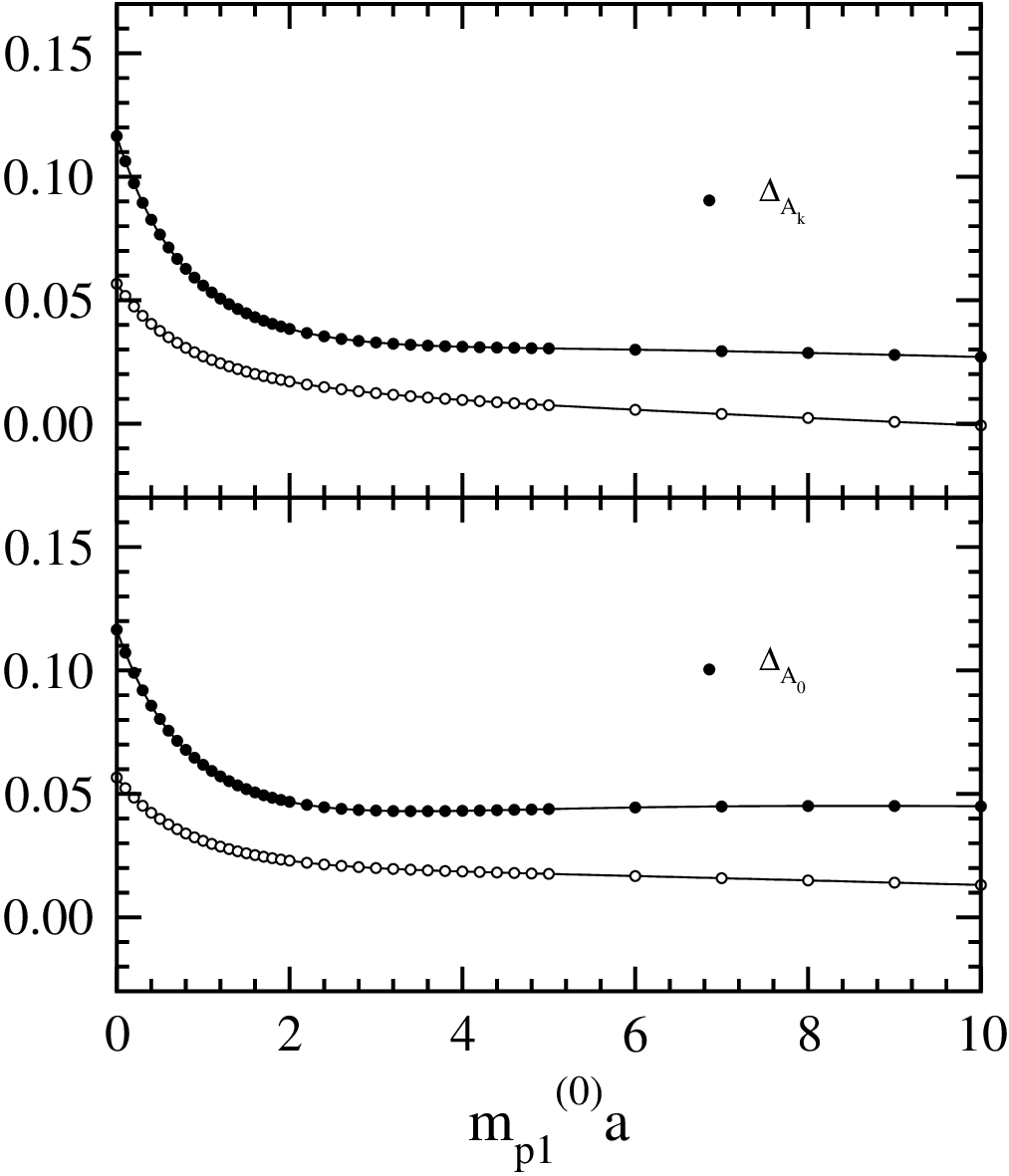}     
}
\caption{$\Delta_{A_\mu}$ for heavy-light case as 
a function of $\mphlo$. Solid symbols denote the plaquette gauge action
and open ones for the Iwasaki gauge action.}
\label{fig:z_a_hl}
\vspace{8mm}
\end{figure}

Numerical results for
$\Delta_{\gamma_k\gamma_5}$, $c_{A_k}^{(+,-,H,L)}$ and 
$\Delta_{\gamma_0\gamma_5}$, $c_{A_0}^{(+,-)}$ are presented
in Figs.~\ref{fig:a_k_hl} and \ref{fig:a_0_hl} respectively, 
in the case of the plaquette and the Iwasaki gauge actions.
We also plot the $\mphlo$ dependence of $\Delta_{A_\mu}$
in Fig.~\ref{fig:z_a_hl}.
The solid lines represent the interpolation expressed as
\ben
\Delta_{\gamma_k\gamma_5}&=&
\left.\Delta_{\gamma_k\gamma_5}\right|_{\mphlo =0}+
\frac{\sum_{i=1}^5 a_{ni}^{k\gamma}\{\mphlo\}^i}
{1+\sum_{i=1}^5 a_{di}^{k\gamma}\{\mphlo\}^i},
\label{eq:fit_d_a_k_hl}\\
c_{A_k}^{+}&=&
\left.c_{A_k}^{+}\right|_{\mphlo =0}+
\frac{\sum_{i=1}^5 a_{ni}^{k+}\{\mphlo\}^i}
{1+\sum_{i=1}^5 a_{di}^{k+}\{\mphlo\}^i},
\label{eq:fit_+_a_k_hl}\\
c_{A_k}^{(-,H,L)}&=&
\frac{\sum_{i=1}^5 a_{ni}^{k(-,H,L)}\{\mphlo\}^i}
{1+\sum_{i=1}^5 a_{di}^{k(-,H,L)}\{\mphlo\}^i},
\label{eq:fit_-_a_k_hl}\\
\Delta_{A_k}&=&
\left.\Delta_{A_k}\right|_{\mphlo =0}+
\frac{\sum_{i=1}^5 a_{di}^{kZ}\{\mphlo\}^i}
{1+\sum_{i=1}^5 a_{di}^{kZ}\{\mphlo\}^i},
\label{eq:fit_z_a_k_hl}\nn\\
\Delta_{\gamma_0\gamma_5}&=&
\left.\Delta_{\gamma_0\gamma_5}\right|_{\mphlo =0}+
\frac{\sum_{i=1}^5 a_{ni}^{0\gamma}\{\mphlo\}^i}
{1+\sum_{i=1}^5 a_{di}^{0\gamma}\{\mphlo\}^i},
\label{eq:fit_d_a_0_hl}\\
c_{A_0}^{+}&=&
\left.c_{A_0}^{+}\right|_{\mphlo =0}+
\frac{\sum_{i=1}^5 a_{ni}^{0+}\{\mphlo\}^i}
{1+\sum_{i=1}^5 a_{di}^{0+}\{\mphlo\}^i},
\label{eq:fit_+_a_0_hl}\\
c_{A_0}^{-}&=&
\frac{\sum_{i=1}^5 a_{ni}^{0-}\{\mphlo\}^i}
{1+\sum_{i=1}^5 a_{di}^{0-}\{\mphlo\}^i},
\label{eq:fit_-_a_0_hl}\\
\Delta_{A_0}&=&
\left.\Delta_{A_0}\right|_{\mphlo =0}+
\frac{\sum_{i=1}^5 a_{ni}^{0Z}\{\mphlo\}^i}
{1+\sum_{i=1}^5 a_{di}^{0Z}\{\mphlo\}^i},
\label{eq:fit_z_a_0_hl}
\een
where the relative errors to the data are 
a few \% over the range $0\le\mphlo\le 10$.
The values of the parameters 
$a_{ni}^{k,0}$ and $a_{di}^{k,0}$ ($i=1,\dots,5$) are listed 
in Table~\ref{tab:fit_a_k_hl} and \ref{tab:fit_a_0_hl}. 
Here we use the constraint that $c_{A_k}^{(-,H,L)}$ and $c_{A_0}^{(-)}$ 
vanish at $\mphlo=0$.

\section{Mean field improvement}

Let us explain the mean-field improvement on the renormalization 
factors of the vector and the axial vector currents 
in eqs.(\ref{eq:z_v}) and (\ref{eq:z_a}).
We first rewritten their expressions as follows: 
\ben
\frac{Z_{V_\mu}^\latt}{Z_{V_\mu}^\cont}
&=&\sqrt{Z_{Q,\latt}^\lo(\mphlomf)}
\sqrt{Z_{q,\latt}^\lo(\mpllomf)}u\left(1-g^2\Delta_{V_\mu}\right.\nn\\
&&\left.+g^2\frac{C_F}{2}T_{\rm MF}
+\frac{1}{2}\frac{g^2}{Z_{Q,\latt}^\lo}\frac{\p Z_{Q,\latt}^\lo}{\p \mphlo} {\Delta \mph}
+\frac{1}{2}\frac{g^2}{Z_{q,\latt}^\lo}\frac{\p Z_{q,\latt}^\lo}{\p \mpllo} {\Delta \mpl}\right),\\
\frac{Z_{A_\mu}^\latt}{Z_{A_\mu}^\cont}
&=&\sqrt{Z_{Q,\latt}^\lo(\mphlomf)}
\sqrt{Z_{q,\latt}^\lo(\mpllomf)}u\left(1-g^2\Delta_{A_\mu}\right.\nn\\
&&\left.+g^2\frac{C_F}{2}T_{\rm MF}
+\frac{1}{2}\frac{g^2}{Z_{Q,\latt}^\lo}\frac{\p Z_{Q,\latt}^\lo}{\p \mphlo} {\Delta \mph}
+\frac{1}{2}\frac{g^2}{Z_{q,\latt}^\lo}\frac{\p Z_{q,\latt}^\lo}{\p \mpllo} {\Delta \mpl}\right),
\een
where ${\tilde m_{p1,2}}^{(0)}$ and $\Delta m_{p1,2}$ are defined 
in Ref.~\cite{param} and
$T_{\rm MF}$ is the one-loop correction to the mean-field factor
defined by
\ben
u = P^{1/4}=1-g^2\frac{C_F}{2} T_{\rm MF}
\een
with $P$ the plaquette.
We find a detailed description on the derivation of $T_\mf$ 
in Sec.~III of Ref.~\cite{dwf_pt_rg}.
We then replace $u$ by $P^{1/4}$ measured by Monte Carlo simulation.

The mean-field improved $\msbar$ 
coupling $g_\msbar^2(\mu )$ at the scale $\mu$ is obtained using the
lattice bare coupling $g_0^2$ and $P$:
\ben
\dfrac{1}{g_{\overline{\rm MS}}^2(\mu )}
&=& \dfrac{P}{g^2_0} + d_g + c_p +\dfrac{22}{16\pi^2} \log (\mu a)
+N_f\left(d_f -\dfrac{4}{48\pi^2} \log (\mu a)\right)
\label{eq:g2_plaq}
\een
with $N_f$ the number of quark flavor. 
The values of $c_{p}$, $d_g$ and $d_f$ for various 
gauge and quark actions are summarized in Ref.~\cite{lambda_dwf}.
For the improved gauge action 
one may use an alternative formula\cite{cppacs}
\ben
\dfrac{1}{g_{\overline{\rm MS}}^2(\mu )}
&=& \dfrac{c_0 P + 8 c_1 R1+ 16c_2 R2 +8 c_3 R3}{g^2_0} \nn \\
& &+ d_g + (c_0\cdot c_p + 8 c_1\cdot c_{R1}+16 c_2\cdot c_{R2}+
8 c_3\cdot c_{R3})
+\dfrac{22}{16\pi^2} \log (\mu a) \nn\\
&&+N_f\left(d_f -\dfrac{4}{48\pi^2} \log (\mu a)\right),
\label{eq:g2_rg}
\een
where
\ben
P  &=& \frac{1}{3}{\rm Tr} U_{pl}    =1 - c_{p} g_0^2 +O(g_0^4),\\ 
R1 &=& \frac{1}{3}{\rm Tr} U_{rtg}    =1 - c_{R1} g_0^2 +O(g_0^4),\\
R2 &=& \frac{1}{3}{\rm Tr} U_{chr}        =1 - c_{R2} g_0^2 +O(g_0^4),\\
R3 &=& \frac{1}{3}{\rm Tr} U_{plg}=1 - c_{R3} g_0^2 +O(g_0^4),
\een
and the measured values are employed for $P$, $R1$, $R2$ and $R3$. 
We also find the values of $c_{R1}$, $c_{R2}$ and $c_{R3}$ for various 
gauge actions in Ref.~\cite{lambda_dwf}.

\section{Conclusion}

In this paper we have determined the $O(a)$ improvement coefficients 
of the vector and the axial vector currents 
in a mass dependent way at the one-loop level.
Our calculation is made employing the relativistic heavy quark action, 
which we have recently proposed, with the various gauge actions.
The results are presented both for the heavy-heavy and the heavy-light cases
as a function of heavy quark mass.

For convenience we have given a brief description   
about the implementation of the mean field improvement 
for the renormalization factors. 
We are now performing a numerical simulation 
for the heavy-heavy and heavy-light
meson systems using the relativistic heavy quark action with  
the $O(a)$ improved vector and
axial vector currents, whose parameters are mean-field improved at the one-loop level. 
This work would reveal to what extent our relativistic heavy quark formulation
is quantitatively efficient to study the heavy quark physics.

\section*{Acknowledgments}
We are grateful to Dr. N.~Yamada for useful discussions.
This work is supported in part by the Grants-in-Aid for
Scientific Research from the Ministry of Education, 
Culture, Sports, Science and Technology.
(Nos. 13135204, 14046202, 15204015, 15540251, 15740165.)

\newpage

\newcommand{\J}[4]{{ #1} {\bf #2} (#3) #4}
\newcommand{\MPL}{Mod.~Phys.~Lett.}
\newcommand{\IJMP}{Int.~J.~Mod.~Phys.}
\newcommand{\NP}{Nucl.~Phys.}
\newcommand{\PL}{Phys.~Lett.}
\newcommand{\PR}{Phys.~Rev.}
\newcommand{\PRL}{Phys.~Rev.~Lett.}
\newcommand{\AP}{Ann.~Phys.}
\newcommand{\CMP}{Commun.~Math.~Phys.}
\newcommand{\PTP}{Prog.~Theor.~Phys.}
\newcommand{\Suppl}{Prog. Theor. Phys. Suppl.}
\bibliography{basename of .bib file}

\newpage

\begin{sidetable}[htb]
\caption{Values of parameters $v_{ni}^{k(\gamma,+,H,Z)}$ 
and $v_{di}^{k(\gamma,+,H,Z)}$  ($i=1,\dots,5$) 
in the interpolation 
of $\Delta_{\gamma_k}, c_{V_k}^{(+,H)}, \Delta_{V_k}$ for heavy-heavy case
with eqs.(\protect{\ref{eq:fit_d_v_k_hh}$-$\ref{eq:fit_z_v_k_hh}}), respectively.}
\label{tab:fit_v_k_hh}
\newcommand{\cc}[1]{\multicolumn{1}{c}{#1}}
\begin{tabular}{llllllllllll}
\hline
& gauge action   & $v_{n1}^{k}$ & $v_{n2}^{k}$ & $v_{n3}^{k}$ & $v_{n4}^{k}$ & $v_{n5}^{k}$ & $v_{d1}^{k}$ & $v_{d2}^{k}$ & $v_{d3}^{k}$ & $v_{d4}^{k}$ & $v_{d5}^{k}$  \\
\hline
                    &plaquette      &  
 0.014684 &
 0.13219 &
 0.89527 &
$-$0.11174 &
 0.31131 &
 7.3765 &
34.893 &
 2.4179 &
14.916 &
 0.070066 
\\
$\Delta_{\gamma_k}$ &Iwasaki        &
 $-$0.015146 &
  1.8778 &
 $-$0.49846 &
 $-$1.5767 &
 $-$0.31618 &
169.33 &
117.23 &
127.59 &
 13.898 &
  0.0027264 
\\
                    &DBW2           &
 0.010395 &
$-$0.047110 &
$-$4.5545 &
 0.89861 &
$-$1.0730 &
63.029 &
44.429 &
 3.2276 &
13.901 &
$-$0.022777  
\\
\hline
                    &plaquette      &
   25.076 &
 $-$181.54 &
 $-$195.34 &
    2.4509 & 
   $-$4.1924 &
22843. &
 7808.2 &
13344. &
 $-$324.79 & 
  260.72 
\\
$c_{V_k}^+$         &Iwasaki        &
    0.94447 &
   $-$8.8867 &
    2.5958 &
   23.741 &
   $-$1.9005 &
 1351.1 &
$-$2317.0 &
  361.82 &
$-$2370.9 &
  187.06 
\\
                    &DBW2           &
 $-$0.0022351 &
 $-$0.00014326 &
 $-$0.034665 &
 $-$0.043163 &
  0.047341 &
 $-$5.2267 &
 20.895 &
$-$17.228 &
 16.897 &
 $-$9.3839 
\\
\hline
                    &plaquette      &
 0.037455 &
 0.14443 &
 0.32618 &
$-$0.029729 &
 0.056502 &
12.703 &
14.177 &
41.911 &
 3.7738 &
 6.8538 
\\
$c_{V_k}^H$         &Iwasaki        &
$-$0.0029594 &
$-$0.011657 &
 0.0030631 &
$-$0.25435 &
$-$0.37900 &
$-$1.3206 &
 9.8642 &
51.400 &
29.745 &
31.961 
\\
                    &DBW2           &
$-$0.088565 &
$-$0.18583 &
$-$0.22715 &
$-$0.019377 &
$-$0.19151 &
 6.4502 &
 6.1322 &
 7.0131 &
 2.7602 &
 4.6325 
\\
\hline
                    &plaquette      &
$-$0.16445 &
$-$7.8793 &
 0.59083 &
$-$0.45059 &
 0.058239 &
42.738 &
36.668 &
 2.0995 &
 3.3521 &
 0.0079208 
\\
$\Delta_{V_k}$      &Iwasaki        &
 $-$0.057093 &
$-$10.029 &
 $-$5.0701 &
 $-$0.97000 &
 $-$0.56750 &
116.04 &
153.83 &
  4.2651 &
 24.187 &
 $-$0.027538 
\\
                    &DBW2           &
$-$0.026515 &
$-$1.6401 &
$-$5.5463 &
$-$1.0198 &
$-$2.0489 &
43.104 &
62.938 &
17.548 &
24.974 &
$-$0.013591
\\
\hline
\end{tabular}
\end{sidetable}

\clearpage

\begin{sidetable}[htb]
\caption{Values of parameters $v_{ni}^{0(\gamma,+,Z)}$ 
and $v_{di}^{0(\gamma,+,Z)}$  ($i=1,\dots,5$) 
in the interpolation 
of $\Delta_{\gamma_0}, c_{V_0}^+, \Delta_{V_0}$ for heavy-heavy case
with eqs.(\protect{\ref{eq:fit_d_v_0_hh}$-$\ref{eq:fit_z_v_0_hh}}), respectively.}
\label{tab:fit_v_0_hh}
\newcommand{\cc}[1]{\multicolumn{1}{c}{#1}}
\begin{tabular}{llllllllllll}
\hline
& gauge action   & $v_{n1}^{0}$ & $v_{n2}^{0}$ & $v_{n3}^{0}$ & $v_{n4}^{0}$ & $v_{n5}^{0}$ & $v_{d1}^{0}$ & $v_{d2}^{0}$ & $v_{d3}^{0}$ & $v_{d4}^{0}$ & $v_{d5}^{0}$  \\
\hline
                    &plaquette      &
 0.00019720 &
 2.7579 &
$-$4.4172 &
$-$0.95205 &
$-$0.047970 &
90.897 &
50.506 &
33.105 &
 0.57994 &
 0.021081 
\\
$\Delta_{\gamma_0}$ &Iwasaki        &
  0.015673 &
 $-$1.1138 &
  5.9122 &
 $-$6.4339 &
 $-$3.4423 &
$-$64.281 &
 99.711 &
 73.380 &
 66.707 &
 $-$0.35366 
\\
                    &DBW2           &
 0.012159 &
$-$0.021725 &
$-$4.5857 &
$-$0.90786 &
$-$0.89745 &
38.421 &
46.673 &
14.201 &
 9.0663 &
$-$0.0099410 
\\
\hline
                    &plaquette      &
 0.0071689 &
$-$0.34738 &
 0.011308 &
 0.042259 &
$-$0.17224 &
15.931 &
$-$6.9566 &
12.093 &
$-$2.5858 &
 4.7195 
\\
$c_{V_0}^+$         &Iwasaki        &
  0.019308 &
 $-$4.4339 &
 $-$1.9827 &
  0.50641 &
 $-$0.19121 &
148.80 &
 57.869 &
 62.367 &
$-$13.529 &
  4.7863 
\\
                    &DBW2           &
$-$0.098707 &
$-$0.67466 &
$-$1.2994 &
$-$0.58528 &
$-$0.047428 &
 9.5348 &
25.681 &
22.818 &
10.537 &
 0.82699 
\\
\hline
                    &plaquette      &
$-$0.16426 &
$-$9.6984 &
$-$3.3551 &
$-$1.0849 &
$-$0.26984 &
56.744 &
44.807 &
 5.8125 &
 6.4396 &
 0.0055241
\\
$\Delta_{V_0}$      &Iwasaki        &
$-$0.077730 &
$-$1.6754 &
$-$6.3923 &
$-$1.1045 &
$-$0.99597 &
21.914 &
79.398 &
 4.6312 &
16.091 &
 0.0065184
\\
                    &DBW2           &
$-$0.026338 &
$-$0.76017 &
$-$3.1685 &
$-$1.5062 &
$-$0.78087 &
18.143 &
29.877 &
13.287 &
 7.5750 &
$-$0.0039958
\\
\hline
\end{tabular}
\end{sidetable}

\clearpage

\begin{sidetable}[htb]
\caption{Values of parameters $v_{ni}^{k(\gamma,+,-,H,L,Z)}$ 
and $v_{di}^{k(\gamma,+,H,L,Z)}$  ($i=1,\dots,5$) 
in the interpolation 
of $\Delta_{\gamma_k}, c_{V_k}^{(+,-,H,L)}, \Delta_{V_k}$ for heavy-light case
with eqs.(\protect{\ref{eq:fit_d_v_k_hl}$-$\ref{eq:fit_z_v_k_hl}}), respectively.}
\label{tab:fit_v_k_hl}
\newcommand{\cc}[1]{\multicolumn{1}{c}{#1}}
\begin{tabular}{llllllllllll}
\hline
& gauge action   & $v_{n1}^{k}$ & $v_{n2}^{k}$ & $v_{n3}^{k}$ & $v_{n4}^{k}$ & $v_{n5}^{k}$ & $v_{d1}^{k}$ & $v_{d2}^{k}$ & $v_{d3}^{k}$ & $v_{d4}^{k}$ & $v_{d5}^{k}$  \\
\hline
                    &plaquette      &
 0.0062511 &
 0.052107 &
 0.55038 &
 0.12282 &
 0.081216 &
 8.3279 &
94.574 &
36.477 &
15.865 &
 2.8063 
\\
$\Delta_{\gamma_k}$ &Iwasaki        &
  0.0031637 &
  0.26026 &
  0.55065 &
 $-$0.20725 &
  0.060822 &
 64.787 &
116.19 &
 $-$8.9718 & 
  0.47615 &
  3.3216 
\\
                    &DBW2           &
 0.0021629 &
 0.025944 &
$-$0.039416 &
 0.14083 &
 0.045331 &
 4.9552 &
$-$3.1981 &
26.268 &
25.127 &
 4.2530 
\\
\hline
                    &plaquette      &
 0.0015687 &
$-$0.011647 &
$-$0.56066 &
 0.12164 &
$-$0.21315 &
14.953 &
39.726 &
29.119 &
 5.8968 &
12.722 
\\
$c_{V_k}^+$         &Iwasaki        &
   0.70482 &
  $-$8.0596 &
  $-$9.4795 &
  $-$0.071982 & 
   0.023359 &
1422.3 &
 920.72 &
 921.97 &
   5.9010 & 
  $-$2.2079 
\\
                    &DBW2           &
 $-$0.0053483 &
  0.030398 &
 $-$0.17270 &
  0.20455 &
  0.077260 &
 $-$4.8436 &
 23.308 &
$-$21.287 &
$-$12.747 &
$-$14.273 
\\
\hline
                    &plaquette      &
 $-$0.0022183 &
  0.091343 &
 $-$0.81318 &
  1.0896 &
  0.025931 & 
$-$23.899 &
120.15 &
188.77 &
242.08 &
 75.209 
\\
$c_{V_k}^-$         &Iwasaki        &
 0.0072349 &
 0.033385 &
 0.090949 &
 0.14881 &
 0.0022120 & 
 3.5137 &
32.527 &
29.132 &
42.756 &
 8.4052 
\\
                    &DBW2           &
 0.062422 &
 0.40457 &
$-$0.18140 &
 0.15221 &
 0.0031810 & 
15.041 &
49.001 &
$-$2.6721 &
11.963 &
 7.7052 
\\
\hline
                    &plaquette      &
   0.0054262 &
  $-$0.11193 &
   0.67635 &
  $-$1.0049 &
  $-$0.028206 &
 $-$25.354 &
 249.89 &
$-$910.86 &
$-$443.70 &
$-$147.73 
\\
$c_{V_k}^H$         &Iwasaki        &
$-$0.00071764 &
 0.00044474 &
 0.020620 &
 0.0011663 &
$-$0.000070314 & 
$-$1.3698 &
16.036 &
 3.7954 &
 3.5247 &
$-$0.088466 
\\
                    &DBW2           &
    0.025927 &
   $-$0.65927 &
    2.7483 &
   $-$9.1542 &
   $-$0.11381 &
 $-$371.59 &
 1369.9 &
$-$3734.8 &
$-$1534.4 &
$-$1256.5 
\\
\hline
                    &plaquette      &
 0.039461 &
 0.16335 &
 0.17557 &
 0.53702 &
 0.31432 &
 7.9009 &
 6.4773 &
24.665 &
19.259 &
 3.7410 
\\
$c_{V_k}^L$         &Iwasaki        &
 0.0035068 &
 0.013396 &
$-$0.0074426 &
 0.092313 &
 0.10210 &
 3.5571 &
$-$2.5954 &
24.803 &
16.574 &
 3.6739 
\\
                    &DBW2           &
 $-$0.10580 &
 $-$2.3541 &
 $-$3.0241 &
 $-$0.75090 &
 $-$0.28242 &
 41.820 &
151.64 &
105.43 &
 34.058 &
  9.0980 
\\
\hline
                    &plaquette      &
$-$0.092742 &
 0.18493 &
$-$1.7660 &
 0.073087 &
$-$0.044986 &
$-$1.2430 &
16.954 &
13.695 &
 1.3841 &
 0.61047
\\
$\Delta_{V_k}$      &Iwasaki        &
 $-$0.043329 &
  2.6134 &
 $-$2.8159 &
  0.29050 &
 $-$0.061514 &
$-$59.887 &
 18.652 &
 40.112 &
 $-$0.18607 &
  1.0233
\\
                    &DBW2           &
   0.021307 &
   2.2956 &
   0.36278 &
  $-$1.9582 &
  $-$0.26512 &
$-$171.10 &
  10.537 &
  37.125 &
  83.567 &
   2.1427
\\
\hline
\end{tabular}
\end{sidetable}

\clearpage

\begin{sidetable}[htb]
\caption{Values of parameters $v_{ni}^{0(\gamma,+,-,Z)}$ 
and $v_{di}^{0(\gamma,+,-,Z)}$  ($i=1,\dots,5$) 
in the interpolation 
of $\Delta_{\gamma_0}, c_{V_0}^{(+,-)}, \Delta_{V_0}$ for heavy-light case
with eqs.(\protect{\ref{eq:fit_d_v_0_hl}$-$\ref{eq:fit_z_v_0_hl}}), respectively.}
\label{tab:fit_v_0_hl}
\newcommand{\cc}[1]{\multicolumn{1}{c}{#1}}
\begin{tabular}{llllllllllll}
\hline
& gauge action   & $v_{n1}^{0}$ & $v_{n2}^{0}$ & $v_{n3}^{0}$ & $v_{n4}^{0}$ & $v_{n5}^{0}$ & $v_{d1}^{0}$ & $v_{d2}^{0}$ & $v_{d3}^{0}$ & $v_{d4}^{0}$ & $v_{d5}^{0}$  \\
\hline
                    &plaquette      &
 0.012802 &
 0.49100 &
 0.33452 &
 0.23128 &
 0.040371 &
38.224 &
37.290 &
48.752 &
16.627 &
 1.3978 
\\
$\Delta_{\gamma_0}$ &Iwasaki        &
 0.0077753 &
 0.061860 &
 0.024810 &
 0.14301 &
 0.069308 &
 6.9955 &
 7.3898 &
13.856 &
22.055 &
 2.9415 
\\
                    &DBW2           &
  0.0063022 &
 $-$0.083313 &
  0.56176 &
 $-$0.17488 &
  0.017591 &
$-$11.399 &
 87.354 &
  7.4716 &
 $-$8.3999 &
  1.1371 
\\
\hline
                    &plaquette      &
 0.0048159 &
$-$0.022115 &
$-$0.65883 &
 0.0040211 & 
$-$0.17926 &
23.777 &
68.425 &
36.502 &
19.322 &
 9.7039 
\\
$c_{V_0}^+$         &Iwasaki        &
 $-$0.0094956 &
  0.12375 &
  0.53923 &
  0.36455 &
  0.29355 &
$-$14.207 &
$-$47.045 &
$-$49.638 &
$-$38.991 &
$-$14.935 
\\
                    &DBW2           &
$-$0.047989 &
$-$0.42990 &
$-$0.37339 &
$-$0.55380 &
$-$0.33072 &
13.176 &
23.031 &
24.888 &
25.931 &
10.361 
\\
\hline
                    &plaquette      &
$-$0.014388 &
 0.00032644 & 
 0.13076 &
 0.046446 &
 0.0015886 &
 7.5778 &
21.438 &
13.630 &
 9.7527 &
 0.74691 
\\
$c_{V_0}^-$         &Iwasaki        &
 0.015770 &
$-$0.085430 &
 0.55703 &
 0.30343 &
 0.022609 &
$-$4.5521 &
32.649 &
32.280 &
25.481 &
 2.2470 
\\
                    &DBW2           &
 0.12368 &
 1.3600 &
$-$0.071861 & 
 0.088198 &
 0.14040 &
24.213 &
35.781 &
 4.4921 &
 1.9685 &
 5.1811 
\\
\hline
                    &plaquette      &
$-$0.092162 &
 0.56122 &
$-$0.53593 &
 0.082763 &
$-$0.054690 &
$-$5.1290 &
$-$0.12544 &
 5.3433 &
$-$1.1662 &
 0.88503
\\
$\Delta_{V_0}$      &Iwasaki        &
    0.18102 &
   44.773 &
 $-$167.94 &
   16.819 &
   $-$4.2353 &
$-$1236.9 &
 3747.4 &
 2472.3 &
   43.654 &
   76.008
\\
                    &DBW2           &
   0.051388 &
   4.0357 &
   2.3373 &
  $-$1.5950 &
  $-$0.26105 &
$-$335.95 &
$-$109.32 &
 $-$56.028 &
  89.233 &
   2.1341
\\
\hline
\end{tabular}
\end{sidetable}

\clearpage

\begin{sidetable}[htb]
\caption{Values of parameters $a_{ni}^{k(\gamma,+,H,Z)}$ 
and $a_{di}^{k(\gamma,+,H,Z)}$  ($i=1,\dots,5$) 
in the interpolation 
of $\Delta_{\gamma_k\gamma_5}, c_{A_k}^{(+,H)}, \Delta_{A_k}$ for heavy-heavy case
with eqs.(\protect{\ref{eq:fit_d_a_k_hh}$-$\ref{eq:fit_z_a_k_hh}}), respectively.}
\label{tab:fit_a_k_hh}
\newcommand{\cc}[1]{\multicolumn{1}{c}{#1}}
\begin{tabular}{llllllllllll}
\hline
& gauge action   & $a_{n1}^{k}$ & $a_{n2}^{k}$ & $a_{n3}^{k}$ & $a_{n4}^{k}$ & $a_{n5}^{k}$ & $a_{d1}^{k}$ & $a_{d2}^{k}$ & $a_{d3}^{k}$ & $a_{d4}^{k}$ & $a_{d5}^{k}$  \\
\hline
                            &plaquette      &
$-$0.019142 &
$-$0.047713 &
$-$0.0062276 &
$-$0.017168 &
 0.035701 &
 1.7330 &
 5.6259 &
 1.2548 &
 4.7507 &
 0.27302 
\\
$\Delta_{\gamma_k\gamma_5}$ &Iwasaki        &
$-$0.011685 &
 0.10626 &
$-$0.35238 &
 0.26829 &
 0.20572 &
$-$7.0320 &
16.561 &
 1.8063 &
22.443 &
 1.3801 
\\
                            &DBW2           &
$-$0.0045598 &
 0.11534 &
 0.022512 &
 0.11025 &
$-$0.00049818 & 
 5.1886 &
 0.97387 &
 4.8579 &
 0.79928 &
$-$0.012343 
\\
\hline
                            &plaquette      &
 0.018067 &
$-$0.14688 &
$-$0.12254 &
$-$0.38779 &
$-$0.20974 &
13.975 &
 9.5264 &
35.820 &
13.366 &
11.765 
\\
$c_{A_k}^+$                 &Iwasaki        &
$-$0.029644 &
 0.010150 &
$-$0.27996 &
$-$0.30399 &
$-$0.12102 &
 0.51603 &
 9.4466 &
18.071 &
11.317 &
 5.4941 
\\
                            &DBW2           &
$-$0.15917 &
$-$1.5545 &
$-$2.5459 &
$-$1.1665 &
$-$0.045035 &
18.164 &
67.405 &
67.406 &
33.382 &
 1.2761 
\\
\hline
                            &plaquette      &
$-$0.021064 &
 0.11720 &
$-$0.0045585 & 
 0.037022 &
 0.025896 &
 7.7376 &
 1.2405 &
 8.1200 &
 0.0044767 &
 2.6000 
\\
$c_{A_k}^H$                 &Iwasaki        &
 0.00045423 &
 0.21074 &
 0.23023 &
 0.10905 &
 0.061551 &
15.102 &
14.308 &
23.396 &
 4.7100 &
 5.0676 
\\
                            &DBW2           &
 0.072677 &
 0.62160 &
 0.53520 &
 0.19668 &
 0.019528 &
17.971 &
38.152 &
24.444 &
10.407 &
 1.0515 
\\
\hline
                            &plaquette      &
$-$0.17057 &
$-$8.9210 &
$-$5.1962 &
$-$0.59753 &
$-$0.30081 &
39.166 &
51.312 &
35.011 &
 1.9229 &
 2.9460
\\
$\Delta_{A_k}$              &Iwasaki        &
$-$0.10682 &
$-$2.6283 &
$-$4.7827 &
$-$0.20855 &
$-$0.23399 &
26.635 &
58.249 &
71.704 &
 0.47394 &
 5.7885
\\
                            &DBW2           &
$-$0.040023 &
$-$0.033273 &
$-$0.28351 &
$-$0.0094526 &
$-$0.010229 &
 3.7131 &
 6.7739 &
20.057 &
$-$1.3448 &
 1.6373
\\
\hline
\end{tabular}
\end{sidetable}

\clearpage

\begin{sidetable}[htb]
\caption{Values of parameters $a_{ni}^{0(\gamma,+,Z)}$ 
and $a_{di}^{0(\gamma,+,Z)}$  ($i=1,\dots,5$) 
in the interpolation 
of $\Delta_{\gamma_0}, c_{A_0}^+, \Delta_{A_0}$ for heavy-heavy case
with eqs.(\protect{\ref{eq:fit_d_a_0_hh}$-$\ref{eq:fit_z_a_0_hh}}), respectively.}
\label{tab:fit_a_0_hh}
\newcommand{\cc}[1]{\multicolumn{1}{c}{#1}}
\begin{tabular}{llllllllllll}
\hline
& gauge action   & $a_{n1}^{0}$ & $a_{n2}^{0}$ & $a_{n3}^{0}$ & $a_{n4}^{0}$ & $a_{n5}^{0}$ & $a_{d1}^{0}$ & $a_{d2}^{0}$ & $a_{d3}^{0}$ & $a_{d4}^{0}$ & $a_{d5}^{0}$  \\
\hline
                            &plaquette      &
  0.0012408 &
 $-$0.027454 &
  0.15598 &
 $-$0.17366 &
  0.55058 &
 $-$7.9684 &
 80.466 &
$-$14.079 &
 41.179 &
  3.5040 
\\
$\Delta_{\gamma_0\gamma_5}$ &Iwasaki        &
$-$0.00032900 &
 0.041856 &
 0.49428 &
$-$0.078257 &
 0.24277 &
12.589 &
29.569 &
$-$4.1770 &
12.693 &
 1.4849 
\\
                            &DBW2           &
  0.0024558 &
  0.10332 &
  2.0082 &
 $-$0.81526 &
  0.40598 &
 13.920 &
 59.897 &
$-$15.834 &
  9.6390 &
  2.3878 
\\
\hline
                            &plaquette      &
 0.052844 &
 0.098447 &
$-$0.095352 &
 0.000049485 & 
 0.00013520 &
13.949 &
 9.2995 &
 7.5620 &
 4.0175 &
$-$0.050684 
\\
$c_{A_0}^+$                 &Iwasaki        &
 $-$0.020399 &
 $-$0.22848 &
  0.42451 &
 $-$1.5113 &
 $-$0.93644 &
 14.179 &
$-$31.091 &
111.08 &
 36.737 &
 58.550 
\\
                            &DBW2           &
 $-$0.21817 &
 $-$4.0728 &
  1.8993 &
 $-$2.2319 &
 $-$0.061938 & 
 33.245 &
 78.786 &
$-$31.755 &
 49.858 &
  1.5508 
\\
\hline
                            &plaquette      &
$-$0.19696 &
$-$8.7519 &
$-$5.3415 &
$-$0.48001 &
$-$0.10054 &
44.563 &
55.001 &
31.792 &
 5.2293 &
 1.3789
\\
$\Delta_{A_0}$              &Iwasaki        &
$-$0.094508 &
$-$2.4786 &
 0.35337 &
$-$0.032386 &
$-$0.00029766 &
28.057 &
21.317 &
 4.6126 &
$-$0.66769 &
 0.11312
\\
                            &DBW2           &
$-$0.035180 &
 0.032360 &
$-$0.029443 &
 0.0065747 &
$-$0.00021265 &
 2.8063 &
$-$2.8195 &
 3.0682 &
$-$0.27495 &
 0.033404
\\
\hline
\end{tabular}
\end{sidetable}

\clearpage

\begin{sidetable}[htb]
\caption{Values of parameters $a_{ni}^{k(\gamma,+,-,H,L,Z)}$ 
and $a_{di}^{k(\gamma,+,H,L,Z)}$  ($i=1,\dots,5$) 
in the interpolation 
of $\Delta_{\gamma_k\gamma_5}, c_{A_k}^{(+,-,H,L)}, \Delta_{A_k}$ for heavy-light case
with eqs.(\protect{\ref{eq:fit_d_a_k_hl}$-$\ref{eq:fit_z_a_k_hl}}), respectively.}
\label{tab:fit_a_k_hl}
\newcommand{\cc}[1]{\multicolumn{1}{c}{#1}}
\begin{tabular}{llllllllllll}
\hline
& gauge action   & $a_{n1}^{k}$ & $a_{n2}^{k}$ & $a_{n3}^{k}$ & $a_{n4}^{k}$ & $a_{n5}^{k}$ & $a_{d1}^{k}$ & $a_{d2}^{k}$ & $a_{d3}^{k}$ & $a_{d4}^{k}$ & $a_{d5}^{k}$  \\
\hline
                            &plaquette      &
 $-$0.0077994 &
 $-$0.41731 &
  1.1354 &
  0.094127 & 
  0.29482 &
 45.381 &
$-$95.269 &
$-$62.935 &
$-$98.936 &
$-$29.162 
\\
$\Delta_{\gamma_k\gamma_5}$ &Iwasaki        &
   0.010623 &
  $-$8.7336 &
  $-$0.14090 &
  $-$1.3650 &
  $-$0.040416 & 
1596.5 &
1554.8 &
1388.6 &
 375.05 &
   9.6218 
\\
                            &DBW2           &
 $-$0.0027159 &
 $-$0.023333 &
  0.11288 &
 $-$0.091528 &
 $-$0.0081287 &
 18.209 &
$-$16.459 &
$-$67.226 &
$-$42.775 &
$-$14.363 
\\
\hline
                            &plaquette      &
  0.0019484 &
 $-$0.12572 &
  0.15951 &
  0.073993 &
  0.052235 &
$-$22.185 &
$-$50.850 &
$-$46.419 &
$-$26.709 &
 $-$6.6887 
\\
$c_{A_k}^+$                 &Iwasaki        &
 $-$0.0085895 &
  0.10146 &
 $-$0.15900 &
  0.32234 &
  0.081455 &
 $-$9.2200 &
$-$11.177 &
 $-$1.5297 &
$-$89.575 &
$-$19.944 
\\
                            &DBW2           &
$-$0.061215 &
 0.080878 &
$-$0.031488 &
 0.019396 &
 0.0023197 &
 6.6362 &
$-$7.8813 &
 1.1771 &
$-$1.5133 &
$-$1.2521 
\\
\hline
                            &plaquette      &
   0.0079671 &
  $-$0.37407 &
   2.8843 &
  $-$5.3964 &
  $-$0.14091 &
  27.423 &
 238.43 &
$-$806.02 &
$-$384.25 &
$-$198.31 
\\
$c_{A_k}^-$                 &Iwasaki        &
$-$0.0015928 &
 0.010270 &
 0.057064 &
 0.0053304 &
$-$0.00012449 &
 1.8033 &
15.634 &
 4.7995 &
 3.5412 &
 0.0063867 
\\
                            &DBW2           &
 0.0036783 &
 0.00027955 &
 0.016457 &
 0.0096355 &
 0.000079022 & 
$-$1.0696 &
 5.2927 &
 1.5712 &
 2.3740 &
 0.31191 
\\
\hline
                            &plaquette      &
 0.0017201 &
$-$0.0031462 &
 0.053022 &
 0.0013268 &
$-$0.000064448 & 
$-$1.1603 &
14.397 &
 4.2428 &
 3.3529 &
$-$0.040156 
\\
$c_{A_k}^H$                 &Iwasaki        &
 0.0010232 &
$-$0.000089135 & 
 0.030173 &
 0.00049781 & 
$-$0.000051947 &
$-$1.2988 &
11.412 &
 2.6448 &
 2.7244 &
$-$0.067972  
\\
                            &DBW2           &
 0.0010612 &
 0.0021013 &
 0.014049 &
$-$0.00012538 & 
$-$0.000040511 &
$-$1.6824 &
 8.7455 &
 0.27197 &
 2.1129 &
$-$0.11394 
\\
\hline
                            &plaquette      &
$-$0.028427 & 
$-$0.12517 & 
$-$0.025615 & 
$-$0.15291 & 
0.010900 & 
10.057 & 
23.991 & 
10.071 & 
34.909 & 
 3.2330  
\\
$c_{A_k}^L$                 &Iwasaki        &
  0.017833 & 
  $-$0.64125 & 
  $-$3.1710 & 
  4.4169 & 
  0.64578 & 
$-$242.31 & 
$-$174.96 & 
 214.38 & 
 450.52 & 
  33.050  
\\
                            &DBW2           &
0.092039 & 
0.66663 & 
$-$0.51025 & 
1.1278 & 
0.22717 & 
16.138 & 
 5.1945 & 
 9.1509 & 
32.237 & 
 4.3118  
\\
\hline
                            &plaquette      &
$-$0.10998 &
 0.30354 &
$-$2.9348 &
 0.12711 &
$-$0.084194 &
$-$1.8757 &
23.675 &
19.521 &
 1.4942 &
 0.75518 
\\
$\Delta_{A_k}$              &Iwasaki        &
  0.12336 &
$-$42.429 &
  3.5897 &
 $-$2.2117 &
 $-$0.054401 & 
773.10 &
590.97 &
$-$12.310 &
 42.384 &
  0.13072 
\\
                            &DBW2           &
 $-$0.11610 &
$-$12.061 &
 $-$3.5918 &
 $-$1.7161 &
  0.0095736 & 
621.73 &
304.15 &
192.74 &
 23.838 &
 $-$0.43073 
\\
\hline
\end{tabular}
\end{sidetable}

\clearpage

\begin{sidetable}[htb]
\caption{Values of parameters $a_{ni}^{0(\gamma,+,-,Z)}$ 
and $a_{di}^{0(\gamma,+,-,Z)}$  ($i=1,\dots,5$) 
in the interpolation 
of $\Delta_{\gamma_0\gamma_5}, c_{A_0}^{(+,-)}, \Delta_{A_0}$ for heavy-light case
with eqs.(\protect{\ref{eq:fit_d_a_0_hl}$-$\ref{eq:fit_z_a_0_hl}}), respectively.}
\label{tab:fit_a_0_hl}
\newcommand{\cc}[1]{\multicolumn{1}{c}{#1}}
\begin{tabular}{llllllllllll}
\hline
& gauge action   & $a_{n1}^{0}$ & $a_{n2}^{0}$ & $a_{n3}^{0}$ & $a_{n4}^{0}$ & $a_{n5}^{0}$ & $a_{d1}^{0}$ & $a_{d2}^{0}$ & $a_{d3}^{0}$ & $a_{d4}^{0}$ & $a_{d5}^{0}$  \\
\hline
                            &plaquette      &
 0.000068956 &
 0.00047409 &
 0.037244 &
$-$0.015986 &
 0.013863 &
 4.6946 &
21.769 &
$-$6.2023 &
 6.2455 &
 0.67199 
\\
$\Delta_{\gamma_0\gamma_5}$ &Iwasaki        &
 $-$0.0032698 &
  0.046507 &
  0.085071 &
 $-$0.098217 &
 $-$0.0091805 &
 54.923 &
$-$14.421 &
$-$17.932 &
$-$10.622 &
 $-$0.41183 
\\
                            &DBW2           &
  0.0016117 &
 $-$0.018932 &
  0.27891 &
 $-$0.16669 &
  0.060548 &
 20.906 &
 58.558 &
$-$19.071 &
  3.4284 &
  4.5131 
\\
\hline
                            &plaquette      &
 0.016913 &
 0.098293 &
 0.13800 &
 0.11672 &
 0.058833 &
 9.2871 &
49.110 &
25.152 &
12.626 &
 1.7409 
\\
$c_{A_0}^+$                 &Iwasaki        &
$-$0.012639 &
 0.12162 &
$-$0.20061 &
$-$0.49574 &
 0.094685 &
$-$8.5494 &
 5.8278 &
52.240 &
44.235 &
10.945 
\\
                            &DBW2           &
$-$0.10453 &
$-$0.42038 &
 0.25130 &
$-$0.44821 &
$-$0.11844 &
10.887 &
 7.7564 &
 2.5891 &
13.146 &
 6.4704 
\\
\hline
                            &plaquette      &
$-$0.011811 &
$-$0.035056 &
$-$0.57156 &
$-$0.17770 &
$-$0.36685 &
 5.9258 &
76.946 &
38.435 &
42.946 &
 8.8915 
\\
$c_{A_0}^-$                 &Iwasaki        &
  0.0071466 &
 $-$0.24087 &
  0.15744 &
 $-$0.21666 &
  0.13313 &
$-$47.112 &
$-$65.362 &
$-$38.220 &
$-$50.492 &
$-$10.153 
\\
                            &DBW2           &
 0.046117 &
 0.28507 &
$-$0.026074 &
 0.24895 &
 0.045710 &
10.757 &
10.777 &
 6.7615 &
11.691 &
 2.7523 
\\
\hline
                            &plaquette      &
$-$0.10088 &
 0.73718 &
$-$4.0136 &
 0.12455 &
$-$0.11617 &
$-$6.4483 &
33.715 &
29.058 &
 2.8742 &
 1.3900 
\\
$\Delta_{A_0}$              &Iwasaki        &
 $-$0.024699 &
 $-$6.5964 &
 $-$0.47808 &
 $-$0.45893 &
 $-$0.018953 &
136.12 &
130.02 &
 14.889 &
 13.667 &
  0.091530 
\\
                            &DBW2           &
 $-$0.011150 &
  1.7357 &
 $-$1.4097 &
 $-$0.44819 &
 $-$0.13594 &
$-$98.330 &
 54.859 &
 32.635 &
 28.748 &
  1.5363 
\\
\hline
\end{tabular}
\end{sidetable}

\clearpage

\newpage 

\newpage

\clearpage

\newpage 

\clearpage

\newpage 

\clearpage

\newpage 

\clearpage

\newpage 

\clearpage

\newpage 

\clearpage

\newpage

\clearpage

\newpage 

\clearpage

\newpage 

\clearpage

\newpage 

\clearpage

\newpage 

\clearpage

\newpage 

\clearpage

\end{document}